\documentclass[prd,preprint,tightenlines,showpacs,preprintnumbers,nofootinbib,eqsecnum,superscriptaddress]{revtex4-1}

 \usepackage[dvips,final]{graphicx}
  \usepackage{amssymb}
   \usepackage{amsmath}
    \usepackage{amsfonts}
     \usepackage{epsfig}
     \usepackage{color}
      \usepackage{bm}

\usepackage[section]{placeins}
\usepackage{sidecap}
\usepackage{multirow}
\usepackage{booktabs}
\usepackage{array}
\usepackage{tabularx}
\usepackage{xcolor}
\usepackage{pstricks}
\def\frac#1#2{{\begingroup #1\endgroup\over #2}}

\newcommand{\bl}{\mbox{\boldmath $l$}}
\newcommand{\bp}{\mbox{\boldmath $p$}}

\newcommand{\bq}{\mbox{\boldmath $q$}}
\newcommand{\bQQ}{\mbox{\boldmath $Q$}}

\newcommand{\bk}{\mbox{\boldmath $k$}}

\newcommand{\Tr}{{\mbox{\rm Tr}}}

\newcommand{\half}{{1\over 2}}

\def\Pom{{\bf I\!P}}
\def\lsim{\mathrel{\rlap{\lower4pt\hbox{\hskip1pt$\sim$}}
		\raise1pt\hbox{$<$}}}         
\def\gsim{\mathrel{\rlap{\lower4pt\hbox{\hskip1pt$\sim$}}
		\raise1pt\hbox{$>$}}}         

\begin{document}

\title{Central exclusive production of scalar and pseudoscalar charmonia 
in the light-front $k_T$-factorization approach}

\author{Izabela Babiarz}
\email{izabela.babiarz@ifj.edu.pl.pl}
\affiliation{Institute of Nuclear Physics, Polish Academy of Sciences, 
ul. Radzikowskiego 152, PL-31-342 Krak{\'o}w, Poland}

\author{Roman Pasechnik}
\email{roman.pasechnik@thep.lu.se}
\affiliation{Department of Astronomy and Theoretical Physics,
Lund University, SE-223 62 Lund, Sweden}

\author{Wolfgang Sch\"afer}
\email{Wolfgang.Schafer@ifj.edu.pl} 
\affiliation{Institute of Nuclear
Physics, Polish Academy of Sciences, ul. Radzikowskiego 152, PL-31-342 
Krak{\'o}w, Poland}

\author{Antoni Szczurek}
\email{antoni.szczurek@ifj.edu.pl}
\affiliation{Institute of Nuclear
Physics, Polish Academy of Sciences, ul. Radzikowskiego 152, PL-31-342 
Krak{\'o}w, Poland}
\affiliation{Faculty of Mathematics and Natural Sciences,
University of Rzesz\'ow, ul. Pigonia 1, PL-35-310 Rzesz\'ow, Poland\vspace{1cm}}

\begin{abstract}
\vspace{0.5cm}
We study exclusive production of scalar $\chi_{c0}\equiv \chi_c(0^{++})$ and 
pseudoscalar $\eta_c$ charmonia states in
proton-proton collisions at the LHC energies.
The amplitudes for $gg \to \chi_{c0}$ as well as for $gg \to \eta_c$ 
mechanisms are derived in the $k_{T}$-factorization approach.
The $p p \to p p \eta_c$ reaction
is discussed for the first time. We have calculated rapidity,
transverse momentum distributions as well as such correlation
observables as the distribution in relative azimuthal angle and 
$(t_1,t_2)$ distributions. The latter two observables are very different
for $\chi_{c0}$ and $\eta_c$ cases.
In contrast to the inclusive production of these mesons
considered very recently in the literature,
in the exclusive case the cross section for $\eta_c$ is much 
lower than that for $\chi_{c0}$ which is due to a special interplay 
of the corresponding vertices and off-diagonal UGDFs used to calculate 
the cross sections. We present the numerical results for the key observables 
in the framework of potential models for 
the light-front quarkonia wave functions.
We also discuss how different are the absorptive corrections for
both considered cases.
\end{abstract}

\maketitle

\section{Introduction}

The central exclusive diffractive processes in proton-proton collisions at high energies have 
attracted recently a lot of attention. These processes lead to very unusual final states. 
For example, in the central exclusive production one produces one or a few particles
at central rapidities which are fully measured. There are no other tracks 
in the detectors. The incoming protons remain intact (in the virtue of ``elastic diffraction'') 
or are excited into small mass hadronic systems, which disappear into the beam pipe.
We consider here simultaneously two such reactions, $ pp\:\rightarrow p\:\chi_{c0}\:\:p$ 
and $pp\:\rightarrow p\:\eta_{c}\:\:p$, which are well suited to be analysed in the framework of 
the so-called Durham model formulated by Khoze, Martin and Ryskin (see Ref.~\cite{KMR} and references 
therein). From the experimental point of view, there is a rapidity 
gap, between each of the protons and the produced $\chi_{c0}$ or $\eta_c$ states. 
These processes hence provide a very clean environment for the study of the produced 
hadronic systems tightly connected to poorly known soft and semi-hard QCD dynamics. 
For a review of conceptual and experimental challenges with such central exclusive production (CEP) 
reactions, see for example Ref.~\cite{Albrow:2010yb}.

The theory of the CEP of single $\chi_{cJ}$, $J=0,1,2$ mesons,
with a correct account for the spin of the mesons and precise kinematics of the production 
process has been worked out earlier by Pasechnik, Szczurek and Teryaev (PST) in a series 
of papers \cite{Pasechnik:2007hm,Pasechnik:2009bq,Pasechnik:2009qc}.
The numerical calculations were done for the Tevatron energies. In this analysis, 
the non-relativistic QCD (NRQCD) methods were applied. So far, only CEP of light pseudoscalar mesons 
was discussed in the literature \cite{Pasechnik:2007hm,LNS2014}. There rather 
nonperturbative effects strongly dominate (see Ref.~\cite{LNS2014}).
Very recently in Ref.~\cite{Machado2020} the production of $\chi_{c0}$ at the LHC
was discussed in the $k_T$-factorisation and saturation dipole-model inspired approaches.
The analysis was performed there in the NRQCD approach and using a single model 
for the unintegrated gluon distribution (UGDs) and a particular prescription 
for the off-diagonal UGD. Given a particular importance of the CEP of heavy quarkonia for ongoing
and future experimental studies, we revisit and extend this analysis to account 
for additional effects and sources for theoretical uncertainties (such as the shapes 
of the charmonia wave functions and a treatment of the absorptive corrections, as well as an accurate treatment of the phase space and production 
kinematics) as well as incorporate the pseudoscalar $\eta_c$ final state for the first time.
\begin{figure}
\includegraphics[width=7cm]{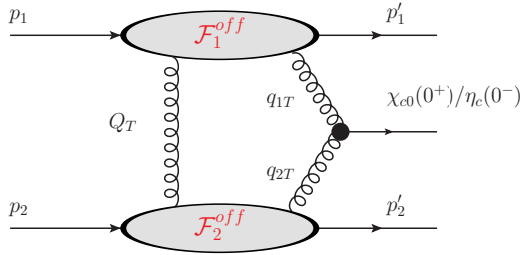}
\caption{Generic diagram for the Durham model approach to the considered
  exclusive production processes.}
\label{fig:diagram_pp_ppM}
\end{figure}

Recently our group showed how to include relativistic corrections
for the inclusive production of $\eta_c$ \cite{BPSS2019} and very recently
for inclusive production of $\chi_{c0}$ \cite{BPSS2020} using the light-cone wave
functions of the charmonia derived from the well-known $c\bar c$ interquark 
potential models. It is the aim of the present paper to do a similar study for 
the exclusive case. In addition, there is no such a study on the 
$p p \to p p \eta_c$ CEP available in the literature.
In contrast, inclusive production of $\eta_c(1S)$ was measured by the LHCb collaboration
in proton-proton collisions for $\sqrt{s}$ = 7, 8, 13 TeV \cite{LHCb}.
Is such a measurement possible for the exclusive production of $\eta_c$?
This study is a first step to address this important question.
An analysis of inclusive diffractive production of $\eta_c(1S)$ was done recently in \cite{Tichouk:2020dut}.

In the present paper, we wish to discuss in parallel the exclusive production process
of both the scalar $\chi_{c0}$ and pseudoscalar $\eta_c$ quarkonia 
For illustration of the corresponding production mechanism initially proposed 
by the Durham group~\cite{KMR}, see Fig.~\ref{fig:diagram_pp_ppM}. 
We start with a brief introduction into the formalism for $pp\to p\, \chi_{c0} \, p$ 
and  $pp\to p\,\eta_c\,p$ reactions based upon the Durham model of CEP \cite{KMR} setting
up the necessary notation and conventions. In this model, a quarkonium state is produced 
via fusion of two virtual active gluons accompanied by an extra exchange with a screening 
gluon as is shown in Fig.~\ref{fig:diagram_pp_ppM}. The additional exchange 
of the gluon provides colour conservation and hence the effective color singlet exchange in the $t$-channel.
As a result, in the final state, in addition to the meson produced mainly at central rapidities, 
there are two forward protons that retain most of their initial energy.
We intend to calculate the integrated cross sections for such processes as well as 
several differential distributions relevant for future measurements. 
We wish to discuss both the hard and soft processes involved in these reactions 
in the light-front QCD approach, to consider several prescriptions on how 
to calculate the off-diagonal UGDs (some of them have already been used 
previously in the literature) and to estimate the absorptive corrections 
in the differential distributions.

\section{Virtual gluon fusion into pseudo(scalar) charmonia}
\label{Sec:LC-vertices}

Below, we shall consider the hard $\chi_{c0}(1P)$ and $\eta_{c}(1S)$ charmonia 
production subprocesses separately.

\subsection{The light-cone amplitude for $g^* g^*\to \chi_{c0}(1P)$ process}

The gluon-gluon fusion vertex is proportional to the reduced amplitude 
${\cal T}_{\mu \nu}$ as follows:
\begin{eqnarray}
{\cal V}^{ab}_{\mu \nu} (g^* g^* \to \chi_{c0}) &=& 4 \pi \alpha_s {\Tr[t^a t^b] 
\over \sqrt{N_c}} \, 2 {\cal T}_{\mu \nu} = {4 \pi \alpha_s \over \sqrt{N_c}} \, 
\delta^{ab} {\cal T}_{\mu \nu} \, ,
\end{eqnarray}
\begin{eqnarray}
 {\cal T}_{\mu \nu}
 = - \delta^\perp_{\mu \nu} (q_1,q_2) G_{\rm TT}(q_1^2,q_2^2)\, 
 + e_\mu^L(q_1) e_\nu^L(q_2) G_{\rm LL}(q_1^2,q_2^2)  \, , 
 \label{eq:gammagamma_amplitude}
\end{eqnarray}
where $\alpha_s$ is the strong coupling, $N_c=3$ and $t^a$ are the number of 
colors and $SU(3)$ group generators in QCD, respectively, and
\begin{eqnarray}
\left( \begin{array}{c} G_{\rm TT} \\ G_{\rm LL} \end{array} \right) &=&  
\begin{pmatrix} - |\bq_1||\bq_2| & (q_1 \cdot q_2) \\  (q_1\cdot q_2) &
  - |\bq_1||\bq_2| \end{pmatrix} \; \left( \begin{array}{c} 
    G_1 \\  G_2 \end{array} \right) \; , \nonumber \\
 \label{eq:FTT_FLL}
\end{eqnarray}
while the relevant kinematical variables are displayed in Fig.~\ref{fig:diagram_pp_ppM}.
Here, we have the projector on transverse polarization states
\begin{eqnarray}
 - \delta^\perp_{\mu \nu} (q_1,q_2) = - g_{\mu \nu} + {1 \over X} 
 \Big( (q_1 \cdot q_2)(q_{1\mu} q_{2\nu} + q_{1 \nu} q_{2\mu}) 
 - q_1^2 q_{2\mu} q_{2\nu}- q_2^2 q_{1\mu} q_{1\nu} \Big) \, ,
\end{eqnarray}
with $X = (q_1 \cdot q_2)^2 - q_1^2 q_2^2$.
The longitudinal polarization vectors read as follows
\begin{eqnarray}
 e_\mu^L(q_1) &=& \sqrt{ {-q_1^2 \over X}} \Big (q_{2\mu} - { q_1 \cdot q_2 \over q_1^2} q_{1\mu} \Big)\, , \qquad
 e_\nu^L(q_2) = \sqrt{ {-q_2^2 \over X}} \Big (q_{1\nu} - { q_1 \cdot q_2 \over q_2^2} q_{2\nu} \Big)\, .
\end{eqnarray}

The convoluted form of reduced amplitude can be written as 
\begin{eqnarray}
  {\cal T}= n^+_\nu n^-_\mu {\cal T_{\mu \nu}} = |\bq_1| |\bq_2| G_{1}(q_1^2,q_2^2) + 
  (\bq_1 \cdot \bq_2) G_{2}(q_1^2,q_2^2)\,,
\end{eqnarray}
in terms of the light cone vectors $n^\pm_\nu = (1,0,0,\pm 1)$.
The form factors here $G_i(\bq_1^2, \bq_2^2)$ have the integral representations in terms of 
the $P$-wave charmonia wave function $\psi_{\chi}(z,\bk)$ (see Ref.~\cite{BPSS2020} for more details)
\begin{eqnarray}
 G_1(\bq_1^2, \bq_2^2) &=& |\bq_1||\bq_2| \,  {4m_c \over \bq_2^2}  \int {dz d^2\bk \over z(1-z) 16 \pi^3} \psi_{\chi}(z,\bk) \,  2z(1-z) (2z-1) \Big[ {1 \over \bl_A^2 + \varepsilon^2} - {1 \over \bl_B^2 + \varepsilon^2}\Big] \nonumber \\
 G_2(\bq_1^2,\bq_2^2) &=&
 4m_c \int {dz d^2\bk \over z(1-z) 16 \pi^3} \psi_{\chi}(z,\bk) \Big[ {1 -z \over \bl_A^2 + \varepsilon^2} + {z \over \bl_B^2 + \varepsilon^2}\Big] \nonumber \\
 &+& {4 m_c \over \bq_2^2} \int {dz d^2\bk \over z(1-z) 16 \pi^3} \psi_{\chi}(z,\bk) 4z(1-z) 
 \Big[ {\bq_2\cdot \bl_A \over \bl_A^2 + \varepsilon^2} - {\bq_2 \cdot \bl_B \over \bl_B^2 + \varepsilon^2}\Big] \, ,
 \label{eq:G1_G2}
\end{eqnarray}
where $z$ is a $c$-quark (or $\bar c$-antiquark) momentum fraction, $\bk$ is the relative $c\bar c$ transverse momentum, $m_c$ is the mass of $c$-quark, and the shorthand notations
\begin{eqnarray}
\varepsilon^2 &=& z(1-z)\bq_1^2 +m_c^2\,,
\qquad \bl_A = \bk - (1-z) \bq_2\,, \qquad 
\bl_B = \bk + z \bq_2 \, ,
\end{eqnarray}
have been introduced.
 
\subsection{The light cone amplitude for $g^*g^*\to \eta_{c}$ process}

In Ref.~\cite{BPSS2019}, we introduced 
the covariant form of the vertex for two off-shell gluon fusion 
into $\eta_{c}$ meson:
\begin{eqnarray}
  {\cal V}^{ab}_{\mu \nu} = (-i) 4\pi \alpha_{s}\epsilon_{\mu\nu\alpha\beta}
  q^{\alpha}q^{\beta}\frac{\delta^{ab}}{2\sqrt{N_c}}2 I(\bq_{1}^{2}, \bq_{2}^{2})\, ,
\end{eqnarray}
where $I(\bq_{1}^{2}, \bq_{2}^{2}) = F_{\gamma^*\gamma^* \to \eta_{c}}(\bq_1^2,\bq_2^2) /(e^2_c \sqrt{N_c}$), or
in the convoluted form: 
\begin{eqnarray}
{\cal V}^{ab} = (-i) 4\pi \alpha_{s}\frac{\delta^{ab}}{\sqrt{N_c}}I(\bq_{1}^{2}, 
\bq_{2}^{2}) |\bq_{1}| |\bq_{2}| \sin(\phi_1 -\phi_2)\, ,  
\end{eqnarray}
with $(\phi_1 - \phi_2)$ being the angle between $\bq_1$ and $\bq_2$.
We then express $I(\bq_1^2,\bq_2^2)$ in terms of light-cone wave functions
as follows \cite{Babiarz:2019sfa}
\begin{eqnarray}
I(\bq_1^2,\bq_2^2) &=&   4 m_c 
\int {dz d^2 \bk \over z(1-z) 16 \pi^3} \psi_{\eta}(z,\bk) 
\Big\{ 
{1-z \over (\bk - (1-z) \bq_2 )^2  + z (1-z) \bq_1^2 + m_c^2}
\nonumber \\
&+& {z \over (\bk + z \bq_2 )^2 + z (1-z) \bq_1^2 + m_c^2}
\Big\} \, ,
\label{eq:FF}
\end{eqnarray}
where $\psi_{\eta}(z,\bk)$ is the wave function of $\eta_c(1S)$ meson.

\section{Matrix element for $p p \to p p M$ reaction}

The amplitude for the CEP process for a given meson $V\equiv \chi_{c0},\,\eta_c$ reads\footnote{Notice a factor 1/2 in the normalization, due to the fact that we use light-cone vectors fulfilling $n^+\cdot n^- = 2$, matching the conventions of PST.}:
\begin{eqnarray}
{\cal M} &=& \frac{s}{2} \pi^2 \frac{1}{2} \frac{\delta_{c_1 c_2}}{N_c^2-1} 
\int d^2\bQQ\; {\cal V}^{c_1c_2}
\frac{{\cal F}_{g}^{\rm off}(x_1, x', \bQQ^2, \bq_{1}^{2}, \mu^2, t_1) 
      {\cal F}_{g}^{\rm off}(x_2, x', \bQQ^2, \bq_{2}^{2},\mu^2 , t_2)}
      { \bQQ^2 \bq_{1}^{2} \bq_{2}^{2}} \, ,
      \label{eq:amplitude}
\end{eqnarray}
in terms of the ``active'' (fusing into $V$) $x_{1,2}$ and ``screening'' $x'$ (connecting
both proton lines) gluon momentum fractions.
The screening gluon carries a transverse momentum $\bQQ$, while the transverse momenta of active gluons are denoted by $\bq_1,\bq_2$. 
The generalized unintegrated gluon distributions (UGDs) also depend on the hard scale of the process $\mu$ (see below).
The $2\to 3$ total cross section can be 
calculated generically as follows:
\begin{eqnarray} \nonumber
 \sigma &=& \frac{1}{2s}\int |{\cal M}|^2 (2\pi)^4 \delta^4(p_1 + p_2 -p_1' -p_2' - p_V) \\
&\times& \Big(\frac{1}{2(2\pi)^3}\Big)^3(dy_1' d^2 \bp_{1}')(dy_2' d^2 \bp_{2}')(dy d^2 \bp_{V})\, ,
\end{eqnarray}
or, following a simplification done in Ref.~\cite{Szczurek:2006bn}, as
\begin{eqnarray}
 \sigma = \frac{1}{2s}\frac{1}{2^8 \pi^4 s} \int |{\cal M}|^2 dt_1 dt_2 dy d\phi \, .
\end{eqnarray}
Above, $t_1 = (p_1 - p_1')^2$, $t_2 = (p_2 - p_2')^2$ and $\phi \in
(0,2\pi)$ is the relative azimuthal angle between the outgoing protons, $s$ is the $pp$
center-of-mass energy squared, $y$ is rapidity of the outgoing meson $V$.

\section{Different approaches to off-diagonal gluon densities}

In the forward limit of small $t_{1,2}\to 0$ corresponding to $\bQQ^2 \simeq \bq_{1,2}^{2}\equiv Q_\perp^2$, 
the generalized UGDs in Eq.~(\ref{eq:amplitude}) are simplified and are considered as functions of only one 
transverse momentum, i.e.
\begin{eqnarray}
{\cal F}_{g}^{\rm off}(x_1, x', \bQQ^2, \bq_{1}^{2}, \mu^2, t_1) 
\to {\cal F}_{g}^{\rm off}(x_1, x', Q_\perp^2, \mu^2, t_1) \, .
\end{eqnarray}
The Khoze-Martin-Ryskin (KMR) prescription for the off-diagonal (``skewed'') UGD includes 
a Sudakov form factor $T_g(q_\perp^2,\mu^2)$ and is typically written as \cite{KMR}
\begin{equation}
{\cal F}_{g, {\rm KMR}}^{\rm off}(x,x',Q_\perp^2,\mu^2; t) =
R_g \frac{d}{{d {\rm ln}}q_{\perp}^2}\Big[xg(x,q_{\perp}^2)
\sqrt{T_g(q_\perp^2,\mu^2)}\Big]_{q_\perp^2 = Q_\perp^2} F(t)  \,,
\label{eq:F_off_KMR}
\end{equation}
with gluon virtualities $q_{\perp}^2 \equiv \bq^2$  playing a role 
of the momentum scale squared in the collinear gluon density $xg(x,q_{\perp}^2)$, 
and with the nucleon form factor $F(t)$ often parameterised in the following two ways
\begin{equation}
    F(t) = \frac{4 m_p^2 - 2.79t}{(4m_p^2 - t)(1-t/0.71)^2}\qquad {\rm or} \qquad
    F(t) = \exp\Big(\frac{bt}{2}\Big)\,, \quad b=4\,{\rm GeV}^{-2}\,,
\end{equation}
with $m_p$ being the proton mass, corresponding to the isoscalar nucleon form factor \cite{Donnachie:1987pu}
or the QCD elastic 
profile factor, respectively. 

The Sudakov form factor is taken as:
\begin{eqnarray}
T_g(q_\perp^2,\mu^2) = \exp \Big[ - \int_{q_\perp^2}^{\mu^2}{\frac{d\bk_\perp^2}{\bk_\perp^2}\frac{\alpha_s(k_\perp^2)}{2 \pi}}\int_0^{1-\Delta}{\Big[ z P_{gg}(z) + \sum_q P_{qg}(z)\Big]dz }
\Big],
\end{eqnarray}
with the hard scale $\mu^2 = M^2_V + q_\perp^2\,$ and $\Delta = k_\perp/(k_\perp + \mu)$.

Regarding the longitudinal momentum fractions, central diffractive production is dominated by the 
region $ x' \ll x_{1,2} \ll 1$.
We thus compute the skewedness correction $R_g$  in Eq.~(\ref{eq:F_off_KMR}) using a method proposed and derived for the collinear off-diagonal 
gluon distributions \cite{Shuvaev:1999ce}:
\begin{eqnarray}
    R_g = \frac{2^{2\lambda+3}}{\sqrt{\pi}}\frac{\Gamma(\lambda+ 5/2)}{\Gamma(\lambda+4)}\,, \qquad
    \lambda = \frac{d}{d{\rm ln}(1/x)}\Big[{\rm ln}\Big(xg(x,q_{\perp}^2)\Big)\Big]\, .
\label{eq:R_g}
\end{eqnarray}

In a slightly off-forward case $t_{1,2}\not=0$, the choice of $Q_{\perp}$ in the off-diagonal 
KMR gluon in Eq.~(\ref{eq:F_off_KMR}) becomes somewhat arbitrary. In practical calculations, 
we use the so-called ``minimum prescription'' proposed by the Durham group, by substituting
$Q^2_{\perp}~ \to {\rm min}(Q^2_{\perp},q^2_{\perp})$ in Eq.~(\ref{eq:F_off_KMR}), 
with transverse momentum of an active gluon $q_{\perp}$ and transverse momentum of 
the screening gluon $Q_{\perp}$. In addition, we suggest a geometrical average 
of active and screening gluon momenta as $Q^2_{\perp} \to \sqrt{Q^2_{\perp} q^2_{\perp}}$ 
-- an option, called BPSS in the following, for brevity.

We vary our results by also using the modified off-diagonal CDHI 
gluon defined as \cite{Cudell:2008gv}
\begin{equation}
{\cal F}^{\rm off}_{g,{\rm CDHI}}(x,x',Q_{\perp},\mu^2; t) = 
R_g \Big[{\partial \over \partial  \log \bar Q^2} 
\sqrt{T_g(\bar Q^2,\mu^2)} \, xg(x,\bar Q^2) \Big] \cdot 
{2 Q_{\perp}^2 q_{\perp}^2 \over Q_{\perp}^4 + q_{\perp}^4} \cdot F(t)\, ,
\label{eq:F_off_CDHI}
\end{equation}
where $\bar Q^2 = (Q^2_{\perp}+q^2_{\perp})/2$. In order to take into account the saturation effects,
we use of the simplest saturation-based UGD inspired by the Golec-Biernat-W\"usthoff (GBW) model \cite{GolecBiernat:1999qd}. In order to extrapolate it into the off-diagonal domain, we use 
the prescription proposed in Ref.~\cite{Pasechnik:2007hm} (further referred to as 
the PST prescription): 
\begin{equation}
    \begin{aligned}
    &{\cal F}^{\rm off}_{g,{\rm GBW}} = \sqrt{Q_{\perp}^2 f^{\rm GBW}_g(x',Q_{\perp}^2)
    q_{\perp}^2 f^{\rm GBW}_g(x,q_{\perp}^2)}\, \sqrt{T_g(q_\perp^2,\mu^2)}\,F(t) \,, \\
    &f^{\rm GBW}_g(x,q_{\perp}^2) = \frac{3\, \sigma_0}{4 \pi^2 \alpha_s} R_0^2 \, q_{\perp}^2 \exp[R_0^2 \, q_{\perp}^2]\,,
\label{eq:F_off_PST}
\end{aligned}
\end{equation}
where $f^{\rm GBW}_g$ is the diagonal GBW UGD, and $x' = |\bQQ|/\sqrt{s}$, 
$R_0 = (x/x_0)^{\lambda/2}$. In practical calculations, we have used the following 
fitted values of the GBW parameters obtained by Golec-Biernat and Sapeta \cite{Golec-Biernat:2017lfv}:
$\sigma_0 = 29.12$ mb, $\lambda = 0.277$, $x_0/10^{-4} = 0.41$, with $\alpha_s(q_\perp^2) = {\rm min}(0.82,\frac{4\pi}{9 \log(Q^2/\Lambda^2_{\rm QCD})})$ and $Q^2 = {{\rm max}}(q_{\perp}^2,0.22 \, 
{\rm GeV}^2)$, $\Lambda_{\rm QCD}^2 = 0.04 \, {\rm GeV}^2$.

As an alternative model based on the color dipole cross section,
we used a fit obtained by Rezaeian and Schmidt \cite{Rezaeian:2013tka}. 
While the GBW model resembles the eikonal unitarization, the 
Rezaeian-Schmidt cross section uses a form of the dipole cross section proposed in Ref.~\cite{Iancu:2003ge} which is motivated by the BFKL
equation and its nonlinear generalizations.
We computed the corresponding UGD $f^{\rm RS}_g(x,|\bq|)$ as
\begin{eqnarray}
f^{RS}_g(x,|\bq|) = \bq^2 {\sigma_0 \over \alpha_s} {N_c \over 8 \pi^2} \int_0^\infty  r dr \,  J_0( |\bq|r) \Big( 1 - {\sigma(x,r) \over \sigma_0}  \Big) \, .  
\end{eqnarray}
As an example, we have used the first set of parameters from Table~I in Ref.~{\cite{Rezaeian:2013tka}}.
\section{Numerical results}

\begin{figure}
    \centering
  \includegraphics[width = 0.7\textwidth]{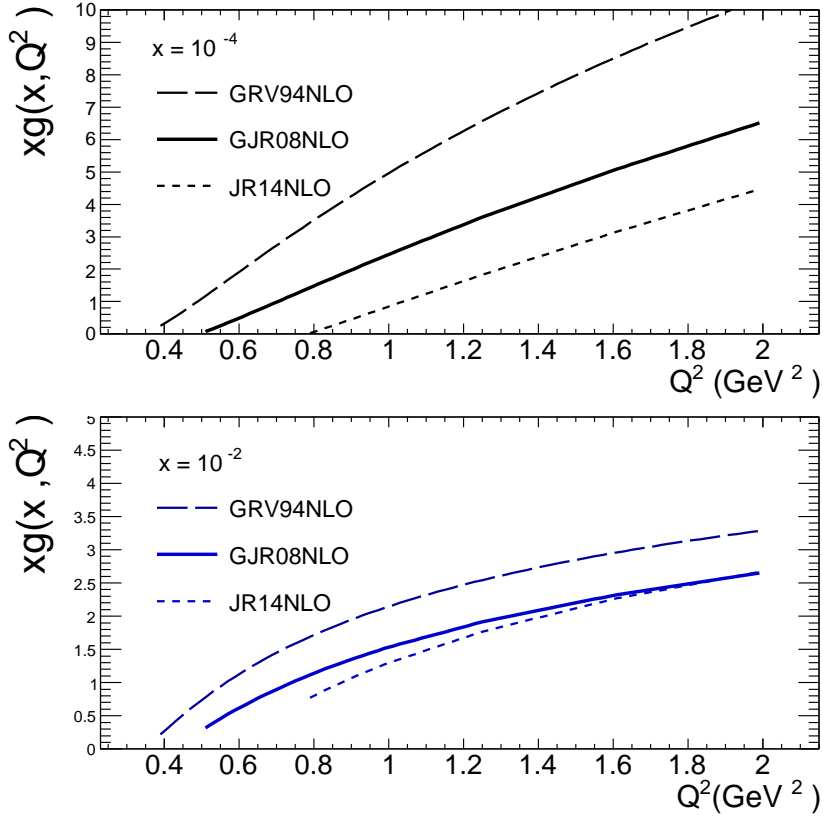}
    \caption{Collinear gluon PDF as a function of the hard scale of the
      process and for typical longitudinal gluon momentum fractions, 
      $x = 10^{-4}$ (upper plot) and $x=10^{-2}$ (lower plot).}
    \label{fig:xgluon}
\end{figure}

In the CEP processes at high energies, it is mandatory to consider gluons 
carrying very small longitudinal momentum fractions $x$. For this purpose, 
in practical calculations we use a few parton distribution functions (PDFs):
JR14NLO \cite{JR14} ($Q_{0T}^{2} = 0.8\, {\rm GeV}^2$), 
GJR08NLO \cite{GJR08} ($Q_{0T}^{2} = 0.5\, {\rm GeV}^2$) and
GRV94NLO \cite{GRV} ($Q_{0T}^{2} = 0.4\, {\rm GeV}^2$). 
In Fig.~\ref{fig:xgluon}, we illustrate the shape of the corresponding 
gluon PDFs at scales and longitudinal momenta fractions 
typical for the considered $p p \to p p \eta_c$ and 
$p p \to p p \chi_{c,0}$ CEP processes. In the range of scales under discussion,
the gluon PDFs from the literature differ considerably.
We do not employ the Durham or CTEQ PDFs for which the initial scales for
evolution are rather high making them difficult to be applied in the context
of the exclusive reactions discussed here.

The total cross sections computed over the full phase space for each PDF mentioned 
above are listed in Tables~\ref{tab:chic0} and \ref{tab:etac}, for $\chi_{c0}$ 
and $\eta_{c}$, respectively. The integrated cross section for $p p \to p p \chi_{c0}$ 
at $\sqrt{s}=$ 13 TeV is shown in Table~\ref{tab:chic0} for different off-diagonal 
UGD prescriptions for the effective $Q_{iT}^2$ summarized as follows:
\begin{itemize}
\item (a) Durham prescription (Eq.~(\ref{eq:F_off_KMR})):  $Q_{iT}^2 = \min (Q_T^2,q_{iT}^2)$\, ,
\item (b) BPSS prescription (Eq.~(\ref{eq:F_off_KMR})):  $Q_{iT}^2 = \sqrt{Q_T^2 q_{iT}^2}$\, ,
\item (c) CDHI prescription (Eq.~(\ref{eq:F_off_CDHI})):  $Q_{iT}^2 = (Q_T^2+q_{iT}^2)/2$\, ,
\item (d) PST off-diagonal UGD (Eq.~(\ref{eq:F_off_PST}))\, .
\end{itemize}
For $\chi_c$ production, these prescriptions lead to similar cross sections of the order of 1 $\mu$b before
including absorption effects. The corresponding gap survival factor is of the order of 0.1 
as will be discussed at the end of this section.
\begin{table}[!h]
    \caption{Total cross section for $\chi_{c0}$ at $\sqrt{s} = 13\, {\rm TeV}$ with $R_g = 1.0$ and $R_{g}$ 
    according to Eq.~(\ref{eq:R_g}). In order to obtain the cross section, several gluon distributions 
    were used with $Q_{0T}^2 \geq 0.4\, {\rm GeV}^2$ for GRV94NLO, $Q_{0T}^2 \geq 0.5\, {\rm GeV}^2$ for GJR08NLO, 
    and $Q_{0T}^2 \geq 0.8\, {\rm GeV}^2$ for JR14NLO. The light-cone form factor for the $gg\to \chi_{c0}$ 
    coupling was calculated using the Buchm\"uller-Tye potential (for more details, see Ref.~\cite{BPSS2020})
    No gap survival factor is included here.}
    \centering
    \begin{tabular}{l|c|c}
    \hline 
    \hline
    KMR Skewed gluon $ 0.8\, {\rm GeV}^{2} \leq Q_{0T}^2$, JR14NLO &
    $\sigma_{\rm tot}$ [nb], $R_g = 1.0$ & $\sigma_{\rm tot}$ [nb], $R_g(x,Q_{iT}^{2})$\\
    \hline
    CDHI, $Q_{iT}^{2} = (Q_{T}^{2}+q_{iT}^{2})/2.$         & $0.42 \cdot10^{3} $ &$ 1.1 \cdot10^{3}$\\
    KMR, $Q_{iT}^{2} = \sqrt{Q_{T}^{2}\cdot q_{iT}^{2}}$   & $0.36 \cdot10^{3} $ &$ 0.94 \cdot10^{3}$\\
    KMR, $Q_{iT}^{2} = \min(Q_{T}^{2},q_{iT}^{2})$          & $0.20 \cdot10^{3}$   & $ 0.52 \cdot10^{3}$ \\
     \hline  
     KMR  Skewed gluon $0.5\, {\rm GeV}^{2} \leq Q_{0t}^2 $, GJR08NLO & $\sigma_{\rm tot}$ [nb], $R_g = 1.0$ &$\sigma_{\rm tot}$ [nb], $R_g(x,Q_{iT}^{2})$\\
    \hline
    CDHI, $Q_{iT}^{2} = (Q_{T}^{2}+q_{iT}^{2})/2.$         & $0.46 \cdot10^{3} $ & $1.57 \cdot10^{3}$\\
    KMR, $Q_{iT}^{2} = \sqrt{Q_{T}^{2}\cdot q_{iT}^{2}}$ & $0.64 \cdot10^{3}$  & $2.1 \cdot10^{3}$\\
    KMR, $Q_{iT}^{2} = \min(Q_{T}^{2},q_{iT}^{2})$          & $0.34 \cdot10^{3}$ & $ 1.1 \cdot10^{3}$ \\
    \hline
    KMR Skewed gluon $0.4\, {\rm GeV}^{2} \leq Q_{0T}^2$, GRV94NLO&$\sigma_{\rm tot}$ [nb], $R_g = 1.0$ &$\sigma_{\rm tot}$ [nb], $R_g(x,Q_{iT}^{2})$\\
    \hline
    CDHI, $Q_{iT}^{2} = (Q_{T}^{2}+q_{iT}^{2})/2.$         & $ 1.88 \cdot10^{3} $  &$9.02 \cdot10^{3}$ \\
    KMR, $Q_{iT}^{2} = \sqrt{Q_{T}^{2}\cdot q_{iT}^{2}}$ & $ 3.03\cdot10^{3}$  & $13.4 \cdot10^{3}$\\    
    KMR, $Q_{iT}^{2} = \min(Q_{T}^{2},q_{iT}^{2}), 0.4 {\rm GeV}^{2} \leq Q_{0T}^2 $ & $1.4 \cdot 10^{3}$ & $6.1 \cdot 10^{3}$ \\
    KMR, $Q_{iT}^{2} = \min(Q_{T}^{2},q_{iT}^{2}), 0.8 {\rm GeV}^{2} \leq Q_{0T}^2$ & $ 0.75\cdot 10^{3}$ & $3.9 \cdot 10^{3}$ \\
    \hline
    PST Skewed gluon &$\sigma_{\rm tot}$ [nb]& -\\
    \hline
    PST prescription, GBW & $ 0.44 \cdot 10^{3}$ & - \\
    PST prescription, RS  & $ 0.52  \cdot 10^{3}$ & - \\
    \hline
    \hline
    \end{tabular}
    \label{tab:chic0}
\end{table}

In Table~\ref{tab:etac} we present similar results for $\eta_c$
production. The total cross section for the $\eta_c$ production
is 3-4 orders of magnitude smaller than that for $\chi_{c0}$,
i.e. surprisingly small. The cross section for the PST prescription for
off-diagonal gluon is quite similar as for the Durham and CDHI
prescriptions in the case of $\chi_{c0}$, while
the spread in the total cross section for $\eta_c$ is much higher.


\begin{table}[!h]
    \caption{The same as in Table~\ref{tab:chic0} but for the $\eta_c$ meson.
    The light-cone form factor for the $gg\to \eta_{c}(1S)$ 
    coupling was calculated using the power-law potential 
    (for more details, see Ref.~\cite{BPSS2019}).}
    \centering
    \begin{tabular}{l|c|c}
    \hline 
    \hline
    KMR Skewed gluon, $0.8 {\rm GeV}^{2} \leq Q_{0T}^2$, JR14NLO &$\sigma_{\rm tot}$ [nb], $R_g = 1.0$ &$\sigma_{\rm tot}$ [nb], $R_g(x,Q_{iT}^{2})$\\
    \hline
    CDHI, $Q_{iT}^{2} = (Q_{T}^{2}+q_{iT}^{2})/2.$         & $1.1$              & $2.4$ \\
    KMR, $Q_{iT}^{2} = \sqrt{Q_{T}^{2}\cdot q_{iT}^{2}}$   & $0.39$             &$ 1.2$ \\
    KMR, $Q_{iT}^{2} = \min(Q_{T}^{2},q_{iT}^{2})$         & $0.13$             &$ 0.25$ \\
    \hline
    KMR Skewed gluon, $0.5 {\rm GeV}^{2} \leq Q_{0T}^2$, GJR08NLO &$\sigma_{\rm tot}$ [nb], $R_g = 1.0$ &$\sigma_{\rm tot}$ [nb], $R_g(x,Q_{iT}^{2})$\\
    \hline
    CDHI, $Q_{iT}^{2} = (Q_{T}^{2}+q_{iT}^{2})/2.$                                 & $2.2$              & $5.6$ \\
    KMR, $Q_{iT}^{2} = \sqrt{Q_{T}^{2}\cdot q_{iT}^{2}}$                           & $0.52$             &$2.1$ \\
    KMR, $Q_{iT}^{2} = \min(Q_{T}^{2},q_{iT}^{2}),0.5 {\rm GeV}^{2} \leq Q_{0T}^2$ & $0.44$             & $1.3$ \\
    KMR, $Q_{iT}^{2} = \min(Q_{T}^{2},q_{iT}^{2}), 0.8 {\rm GeV}^{2} \leq Q_{0T}^2$ & $0.22$            & $0.45$ \\
    
    \hline
    KMR Skewed gluon, $0.4 {\rm GeV}^{2} \leq Q_{0T}^2$, GRV94NLO &$\sigma_{\rm tot}$ [nb], $R_g = 1.0$ &$\sigma_{\rm tot}$ [nb], $R_g(x,Q_{iT}^{2})$\\
    \hline
    CDHI, $Q_{iT}^{2} = (Q_{T}^{2}+q_{iT}^{2})/2.$                                  & $1.2\cdot10^{2}$ &$7.8 \cdot10^{3}$ \\
    KMR, $Q_{iT}^{2} = \sqrt{Q_{T}^{2}\cdot q_{iT}^{2}}$                            & $2.2           $ &$1.3 \cdot10^{3}$ \\
    KMR, $Q_{iT}^{2} = \min(Q_{T}^{2},q_{iT}^{2}), 0.4 {\rm GeV}^{2} \leq Q_{0T}^2$ & $2.8$            &$1.0 \cdot10^{1}$ \\
    KMR, $Q_{iT}^{2} = \min(Q_{T}^{2},q_{iT}^{2}), 0.8 {\rm GeV}^{2} \leq Q_{0T}^2$ & $1.25$           &$2.9$ \\
    \hline
    PST Skewed gluon  &$\sigma_{\rm tot}$ [nb] & -\\
    \hline
    PST, GBW &  $1.9$ & -\\
    PST, RS  &  $4.1$ & \\
    \hline
    \hline
    \end{tabular}
    \label{tab:etac}
\end{table}

\begin{figure}[!h]
    \centering
    \includegraphics[width = 0.4 \textwidth]{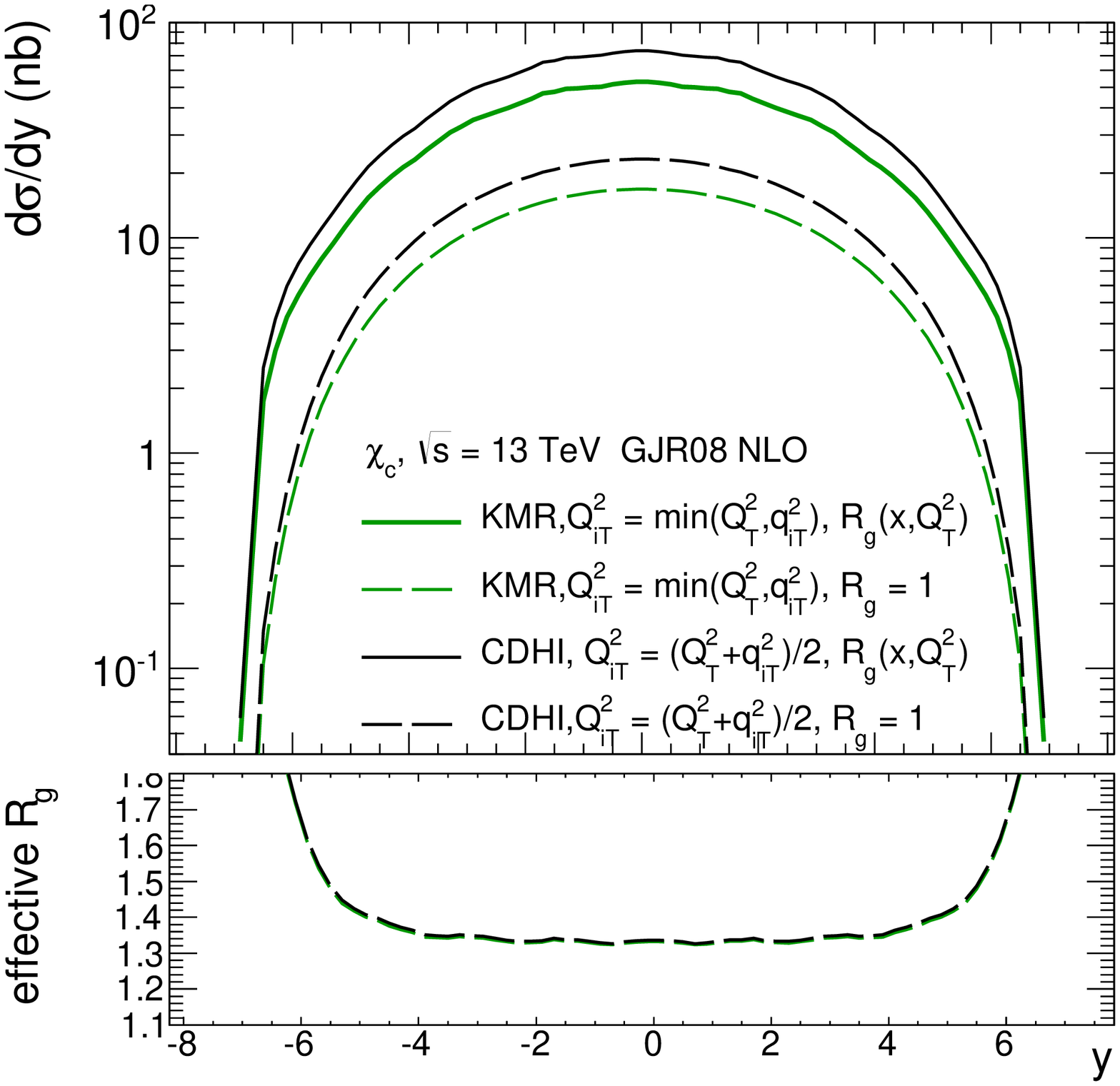}
    \includegraphics[width = 0.4 \textwidth]{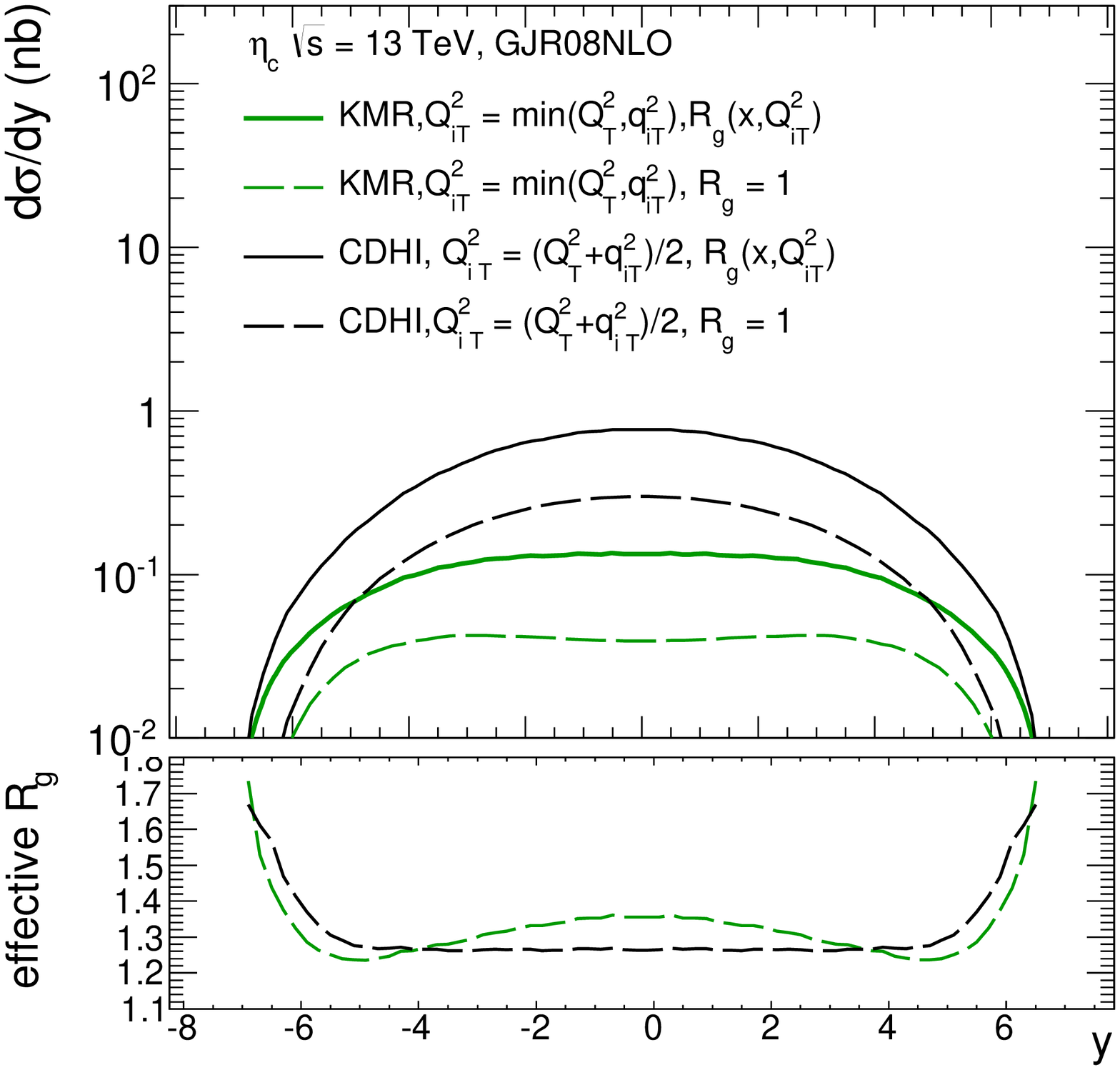}
    \caption{The rapidity distribution for $\chi_{c0}$, $\eta_{c}$ quarkonia CEP and 
effective $R_g$ factor calculated with the GJR08NLO parton distribution 
function. No gap survival factor is included here.}
    \label{fig:dsig_dy_effective_Rg}
\end{figure}
\begin{figure}[!h]
    \centering
    \includegraphics[width = 0.4 \textwidth]{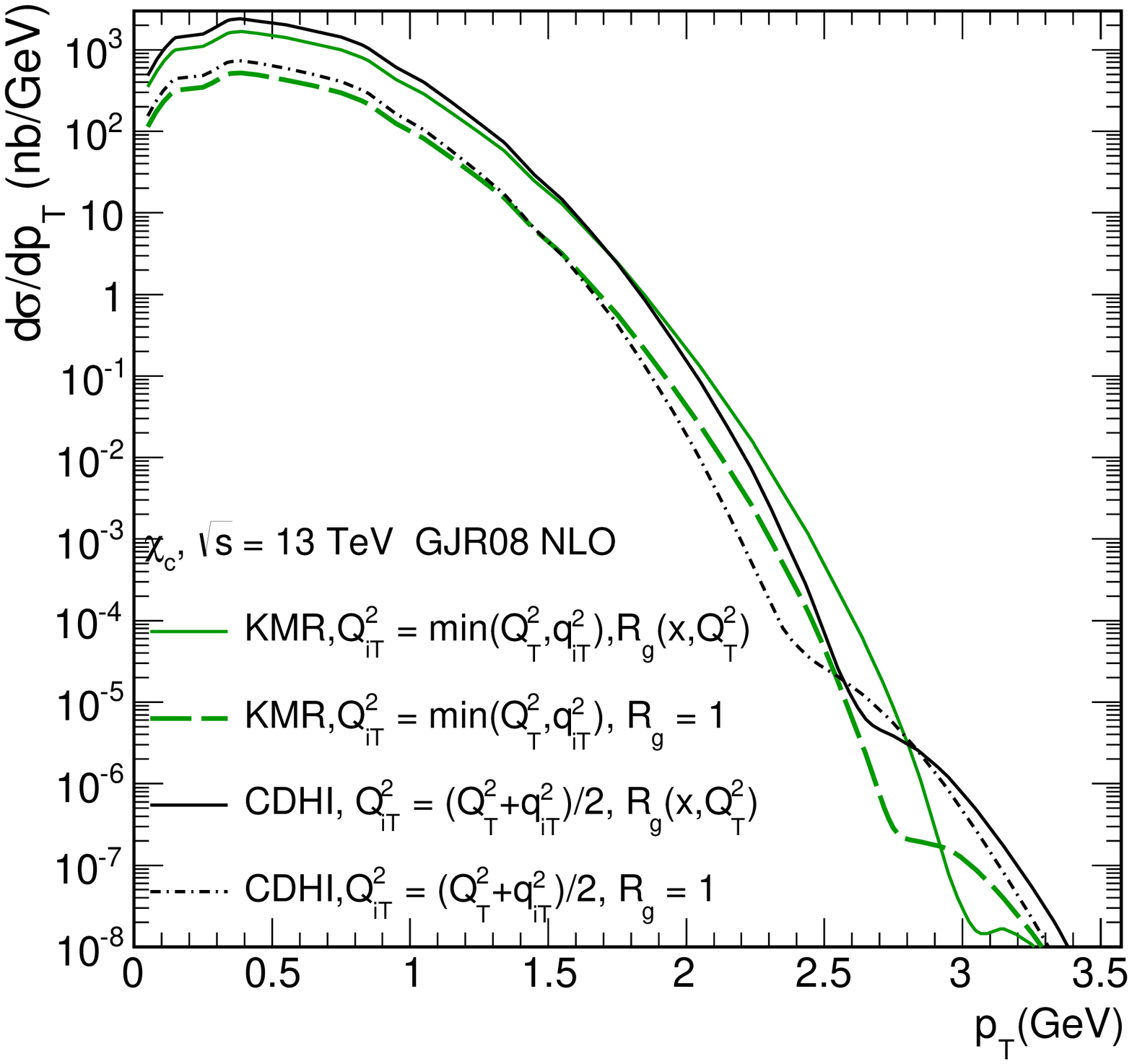}
    \includegraphics[width = 0.4 \textwidth]{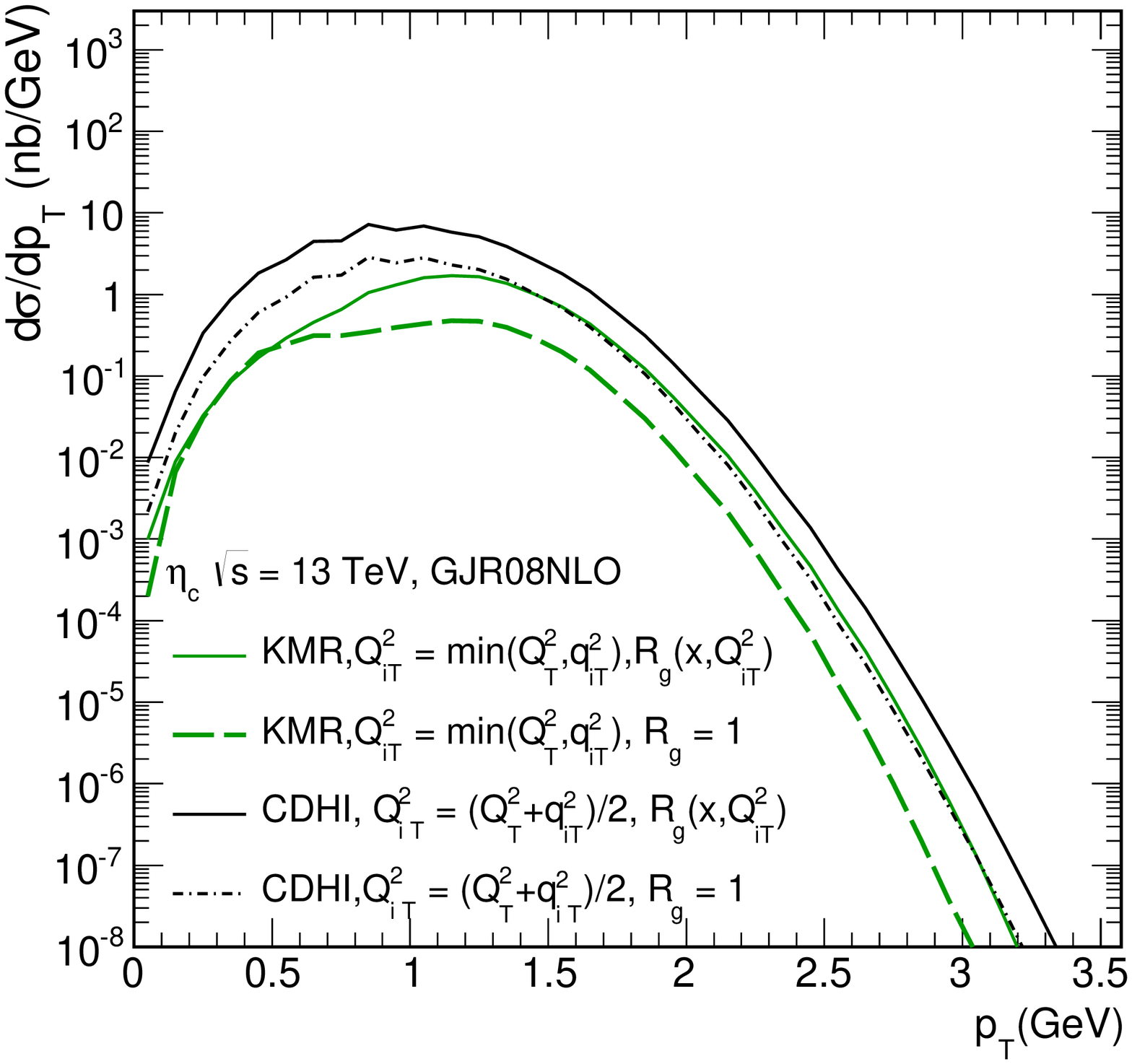}
    \caption{Distribution in transverse momentum of the
      $\chi_{c0}$ (left) and $\eta_{c}$ (right) quarkonia CEP, respectively, with different treatment
      of $R_{g}$ factor. No gap survival factor is included here.}
    \label{fig:dsig_dpt_effective_Rg}
\end{figure}

In Fig.~\ref{fig:dsig_dy_effective_Rg} we show the rapidity distribution of
$\chi_{c,0}$ (left) and $\eta_c$ (right) quarkonia CEP.
We show results for the Durham (min) and CDHI prescription for the off-diagonal
UGDFs. We present results for $R_g =$ 1 as well as with $R_g$ calculated according
to the Shuvaev prescription (see Eq.~(\ref{eq:R_g})). Inclusion of $R_g$ increases
the cross section by a factor of 3-4. While for $\chi_{c0}$ the difference
of the results for the Durham prescription and the CDHI prescription is
small, for $\eta_c$ the difference is of the order of magnitude size.

The distribution in transverse momentum are shown in Fig.~\ref{fig:dsig_dpt_effective_Rg}. 
The distribution for $\eta_c$ and $\chi_{c0}$ CEP are somewhat different. The maximum 
of the cross-section for $\eta_c$ is at $p_T\sim 1\,{\rm GeV}$ and the dip at 
vanishing $p_T$ is more pronounced.

\begin{figure}
    \centering
      \includegraphics[width = 0.4\textwidth]{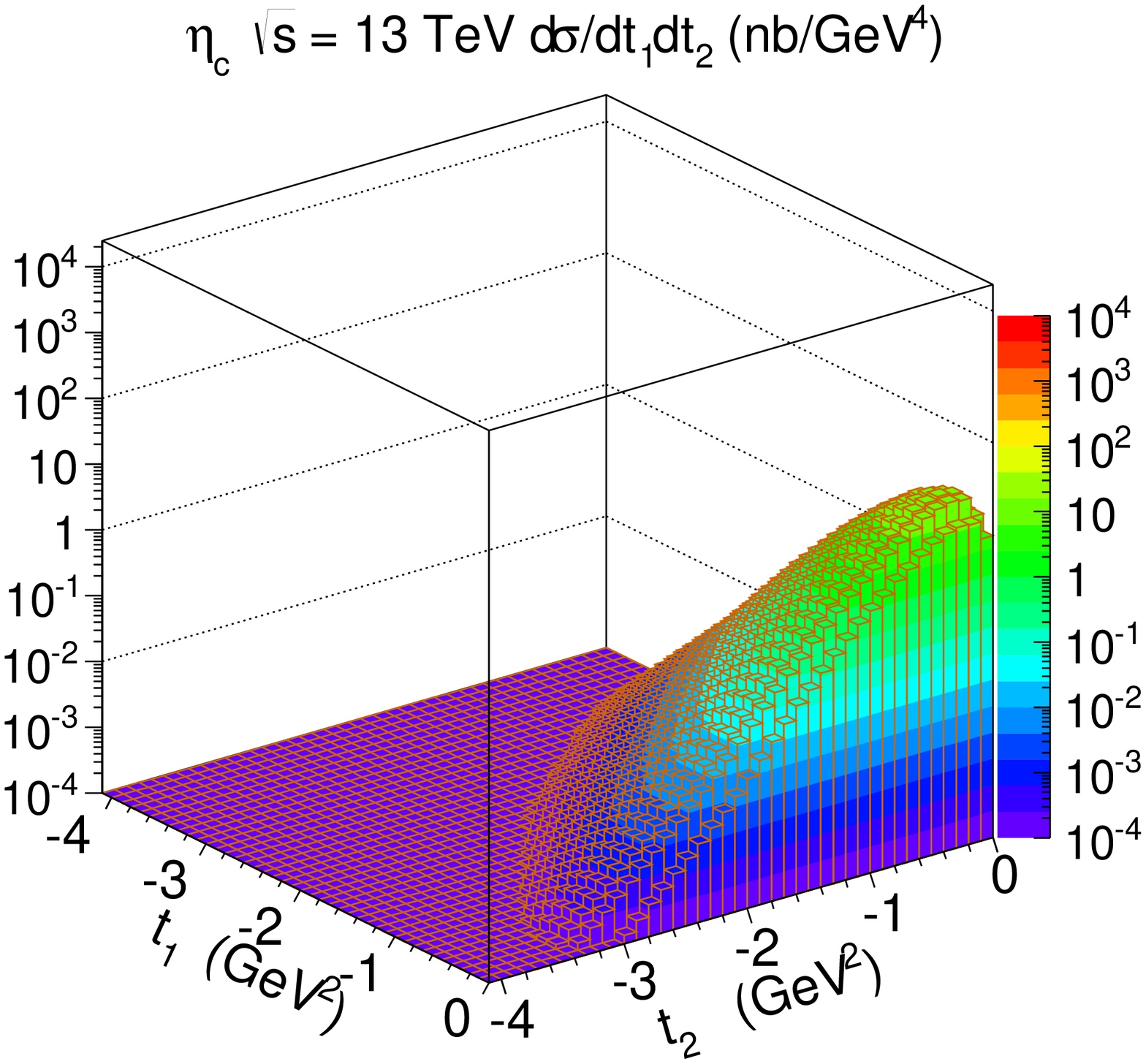}
    \includegraphics[width = 0.4\textwidth]{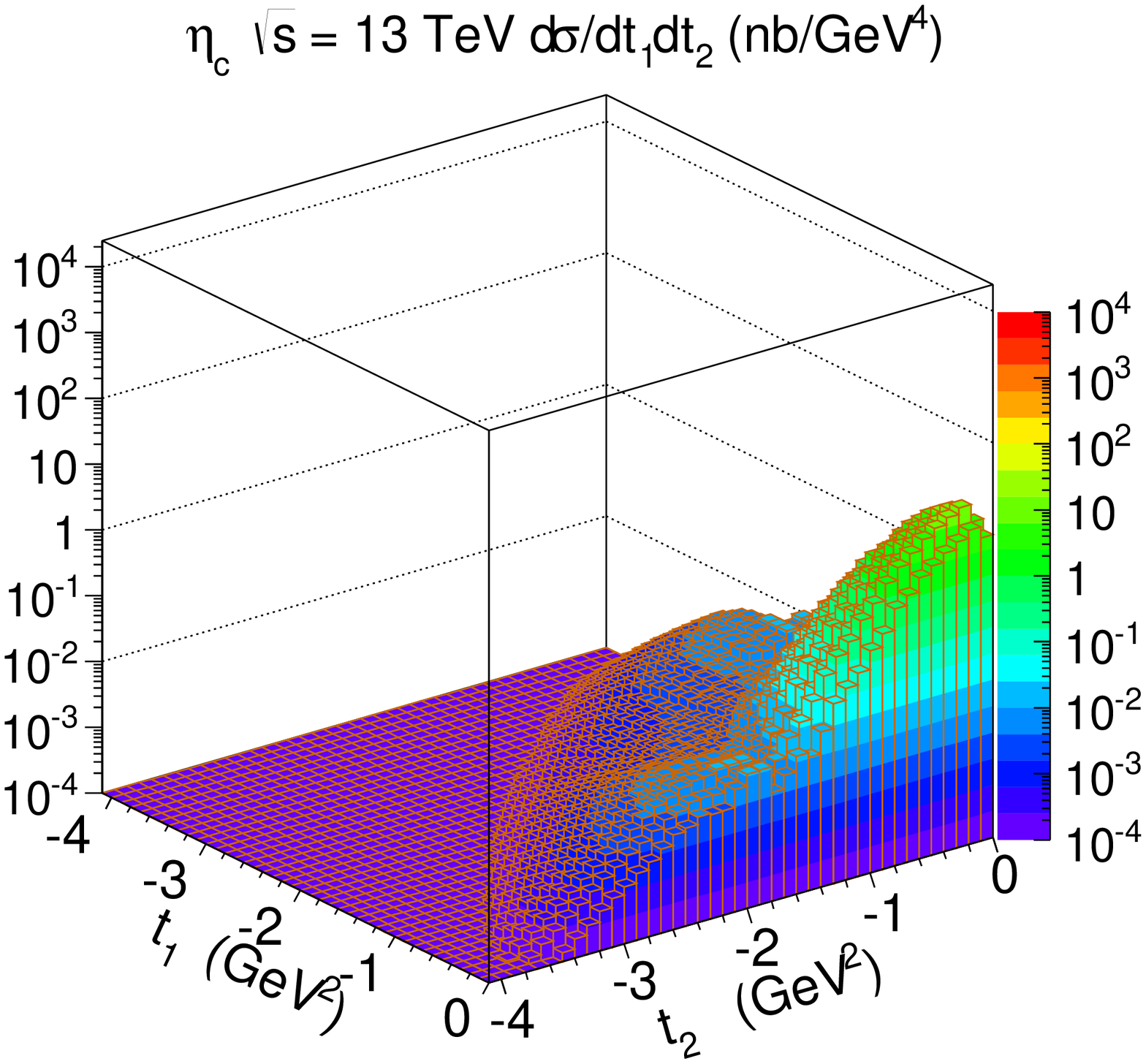}\\
    \includegraphics[width = 0.4\textwidth]{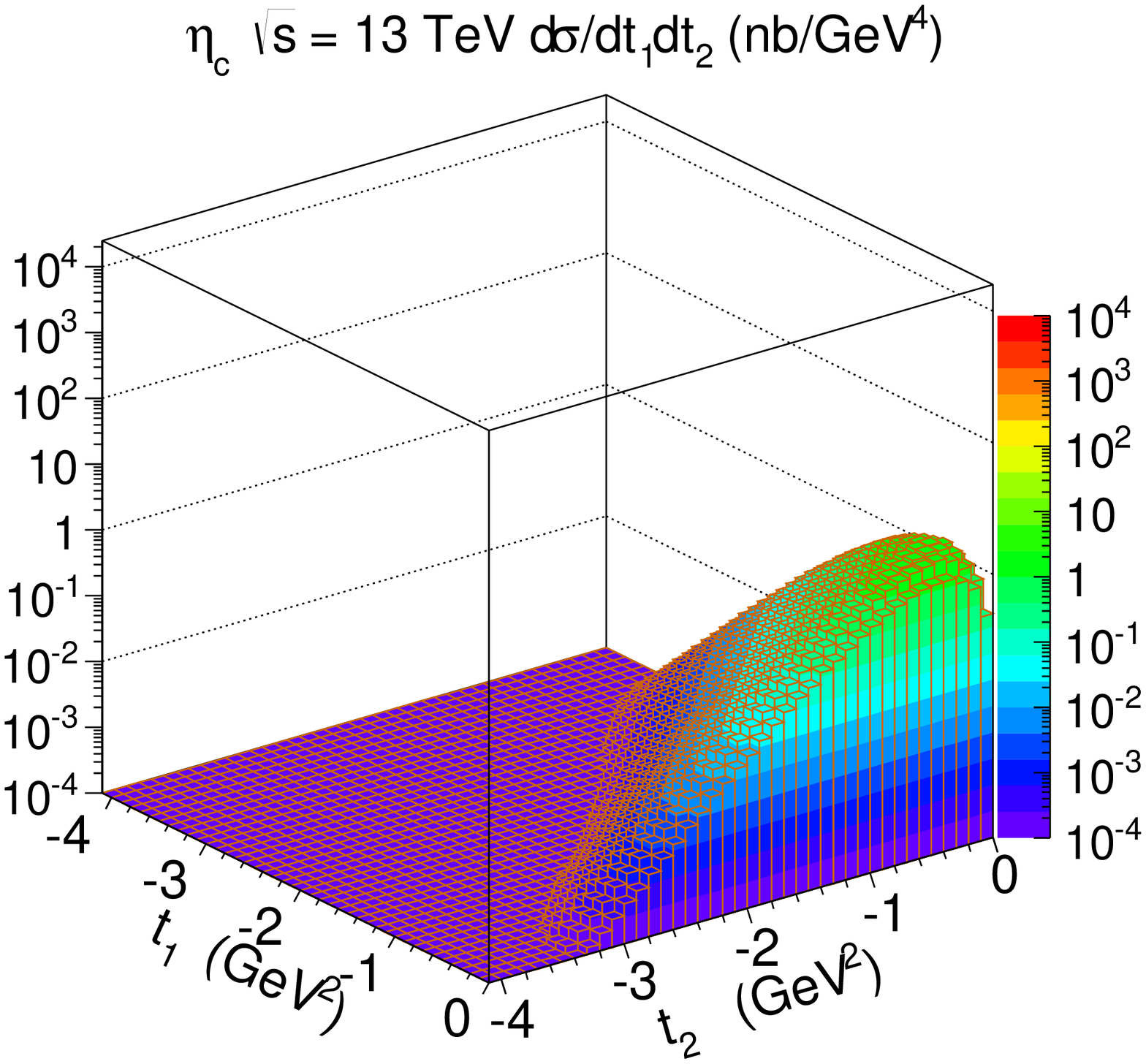}
    \includegraphics[width = 0.4\textwidth]{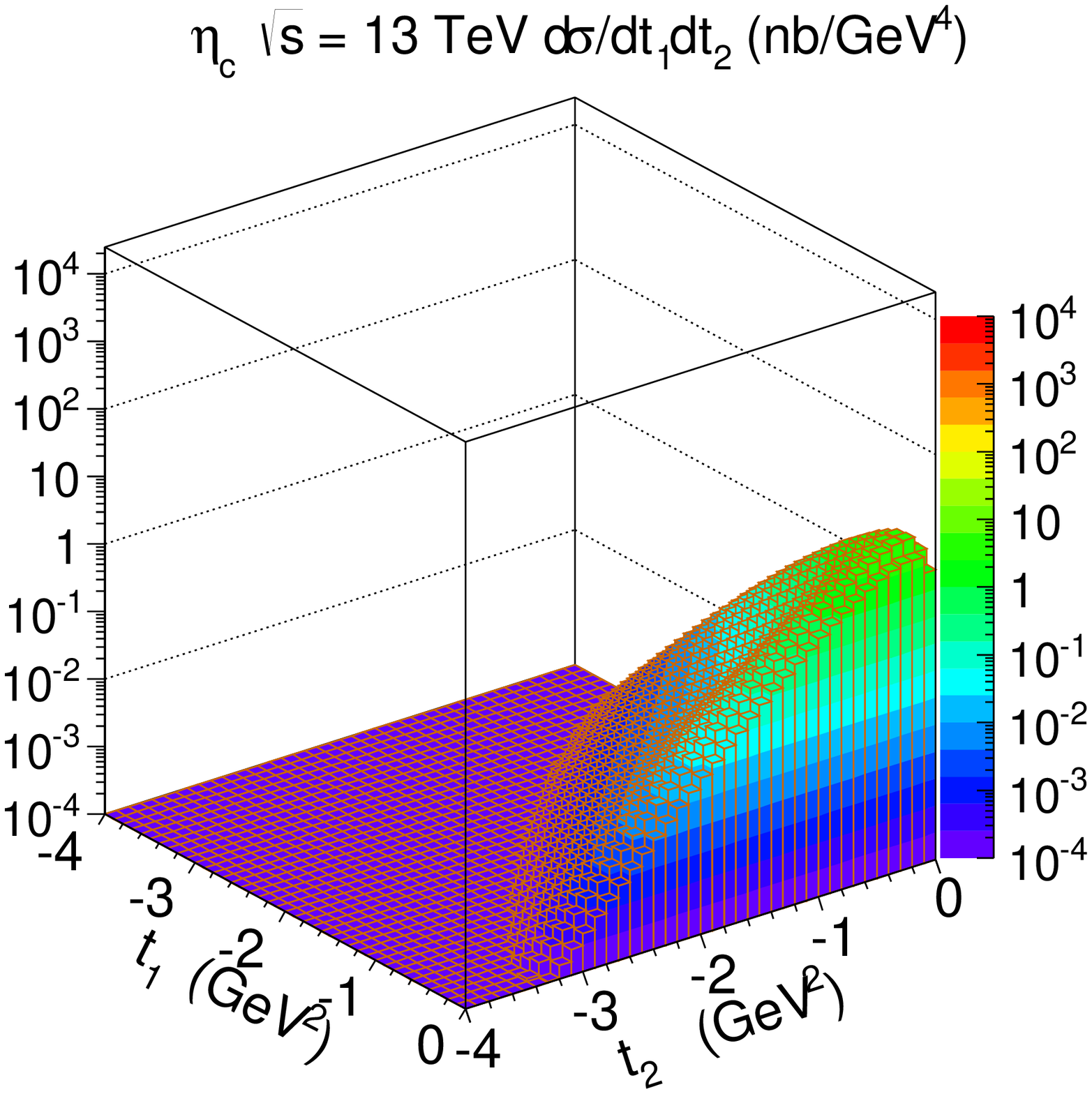}
    \caption{Distribution in $t_{1}\times t_{2}$ for CDHI (left-upper), BPSS (right-upper)
      and Durham (left-lower) prescriptions calculated with the GJR08NLO gluon distribution function and for the PST off-diagonal UGD computed with the diagonal GBW UGD (right-lower) for $\eta_{c}$ CEP for $\sqrt{s} = 13\,{\rm TeV}$.
      No gap survival factor is included here.}
    \label{fig:t1t2_etac}
\end{figure}

In Fig.~\ref{fig:t1t2_etac} we show two-dimensional distributions in
$(t_1,t_2)$ ($t_1$, $t_2$ are four-momenta squared transferred in
the proton lines), for $p p \to p p \eta_c(1S)$ CEP process at 
$\sqrt{s} = 13\,{\rm TeV}$. Here, we present results for several prescriptions 
for the off-diagonal KMR UGDFs: with the Durham prescription -- left-upper panel, the BPSS prescription -- right-upper panel, 
and the CDHI prescription -- left-lower panel as well as the PST prescription
for off-diagonal UGD using the diagonal GBW UGDF -- right-lower panel. No gap survival effect is incorporated here. The results appear to be reasonably stable with respect to a change in the UGDs modelling, while the BPSS prescription differ in the distribution shape.

\begin{figure}
    \centering
    \includegraphics[width = 0.4\textwidth]{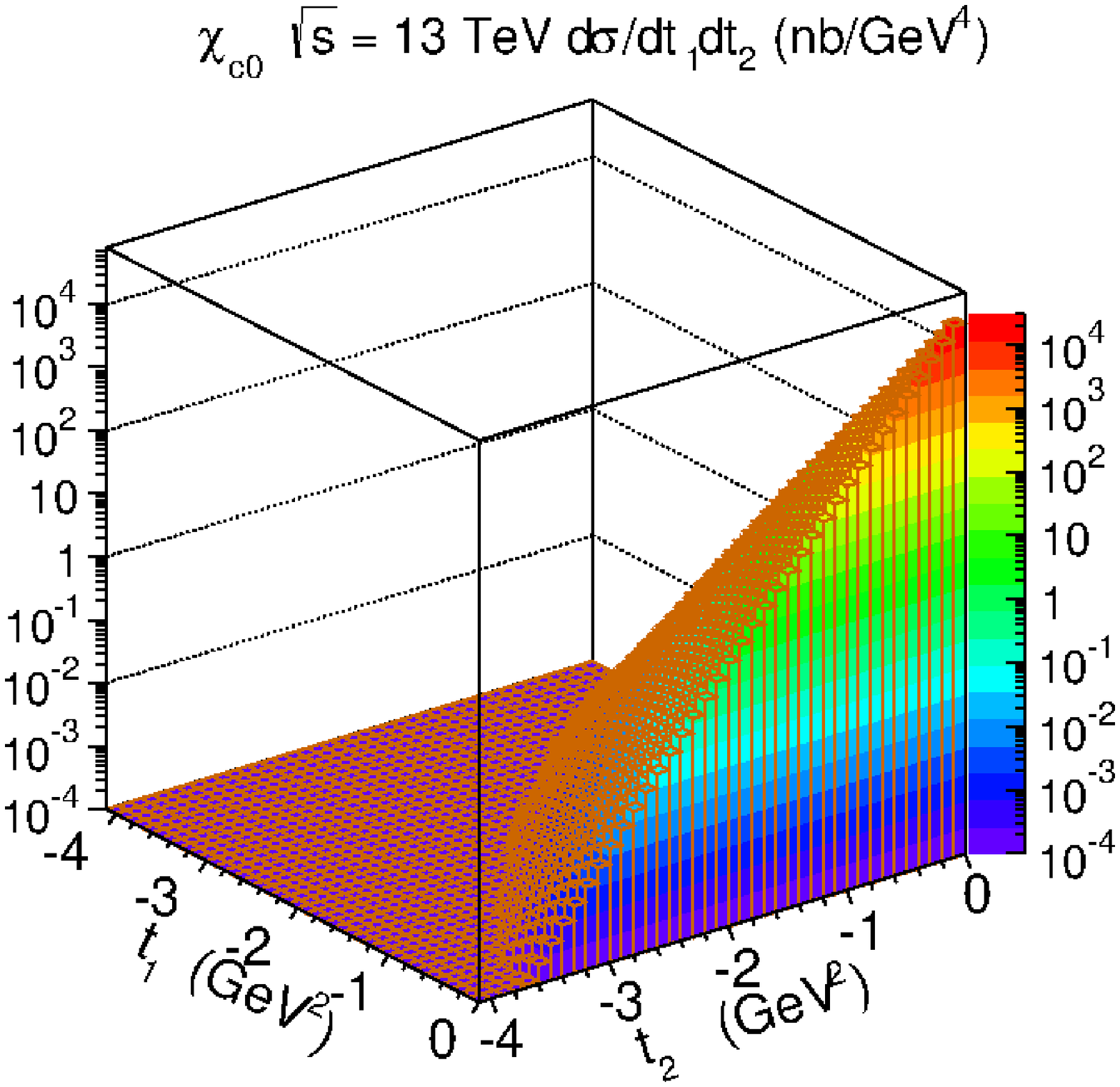}
    \includegraphics[width = 0.4\textwidth]{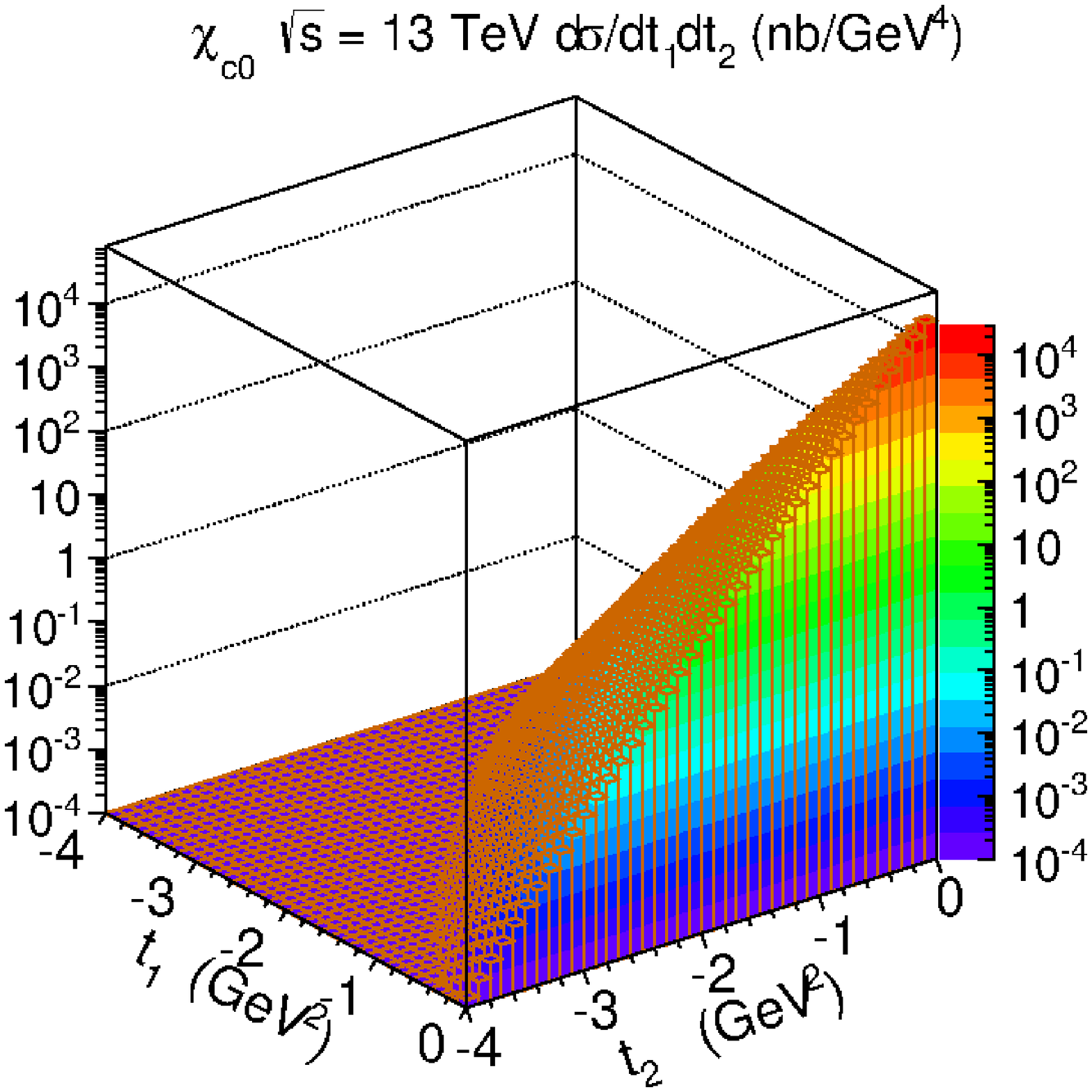}\\
    \includegraphics[width = 0.4\textwidth]{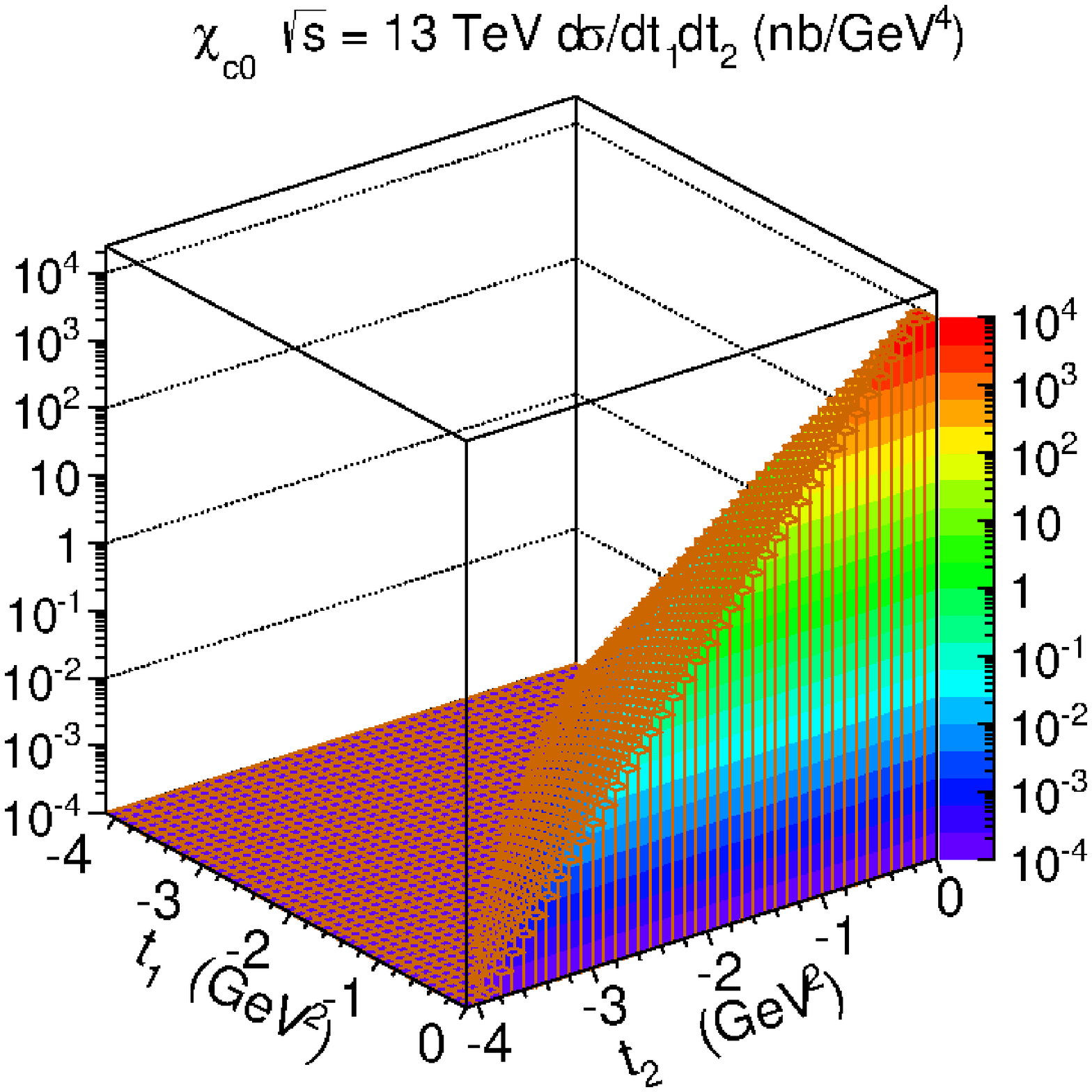}
    \includegraphics[width = 0.4\textwidth]{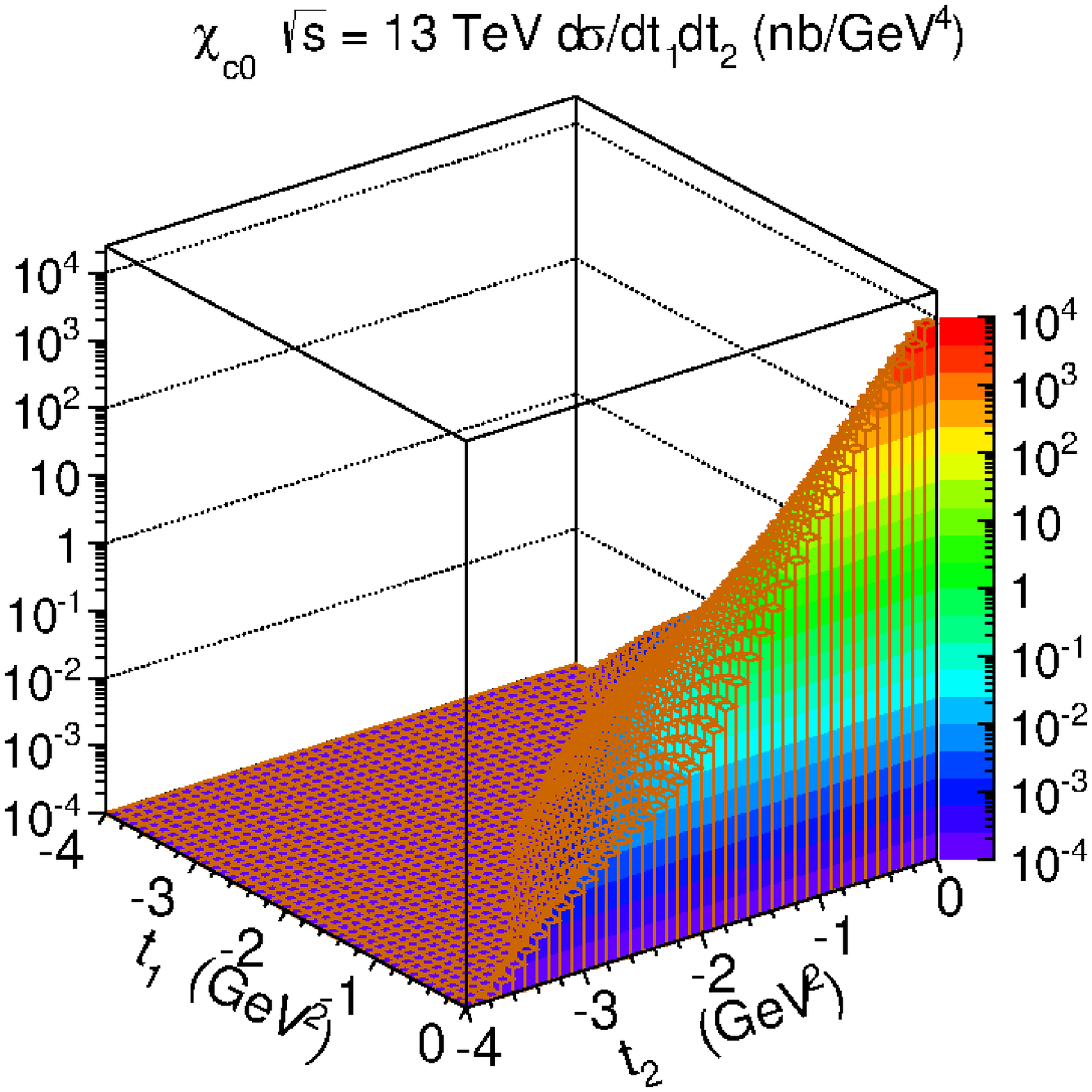}
    \caption{Distribution in $t_{1}\times t_{2}$ for CDHI (left-upper), BPSS 
      ($Q_{iT}^2 = \sqrt{q_{iT}^2,Q_T^2}$)
      (right-upper) and Durham ($Q^2_{iT} = \min(q^2_{iT},Q^2_T)$) (left-lower)
      prescriptions calculated with the GJR08NLO gluon distribution function 
      and for the PST off-diagonal UGD computed with the diagonal GBW UGD (right-lower) 
      for $\chi_{c0}$ for $\sqrt{s} = 13\,{\rm TeV}$.
      No gap survival factor is included here.}
    \label{fig:t1t2_chic}
\end{figure}

For completeness, in Fig.~\ref{fig:t1t2_chic} we show similar results for
the $\chi_{c0}$ production. 
In this case, all prescriptions for effective transverse momenta 
(Durham, BPSS, and CDHI prescriptions) lead to fairly similar results. 
Here the cross sections are peaked at $t_1$ = 0, $t_2$ = 0. 

\begin{figure}
    \centering
    \includegraphics[width = 0.45 \textwidth]{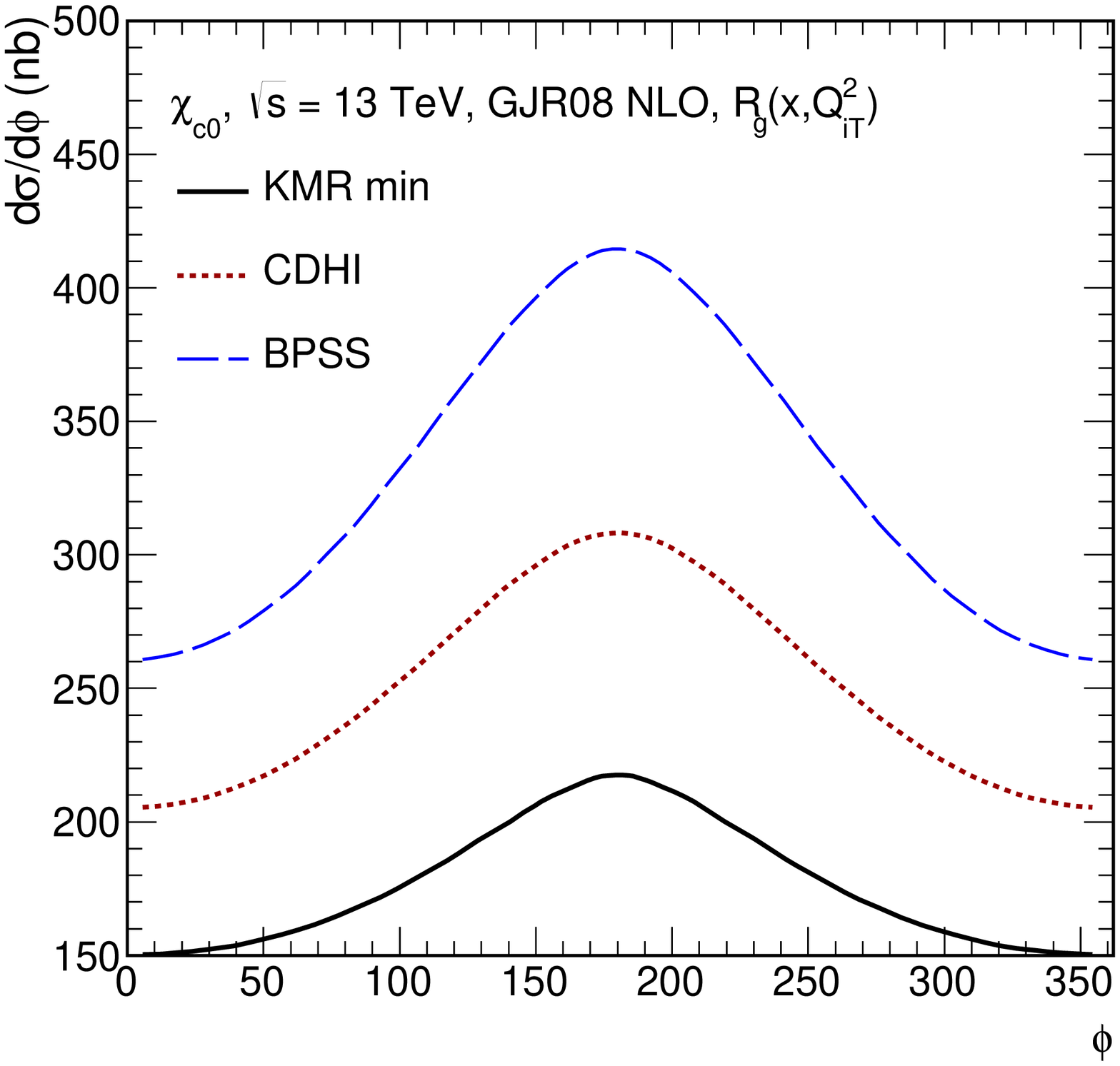}
    \includegraphics[width = 0.45 \textwidth]{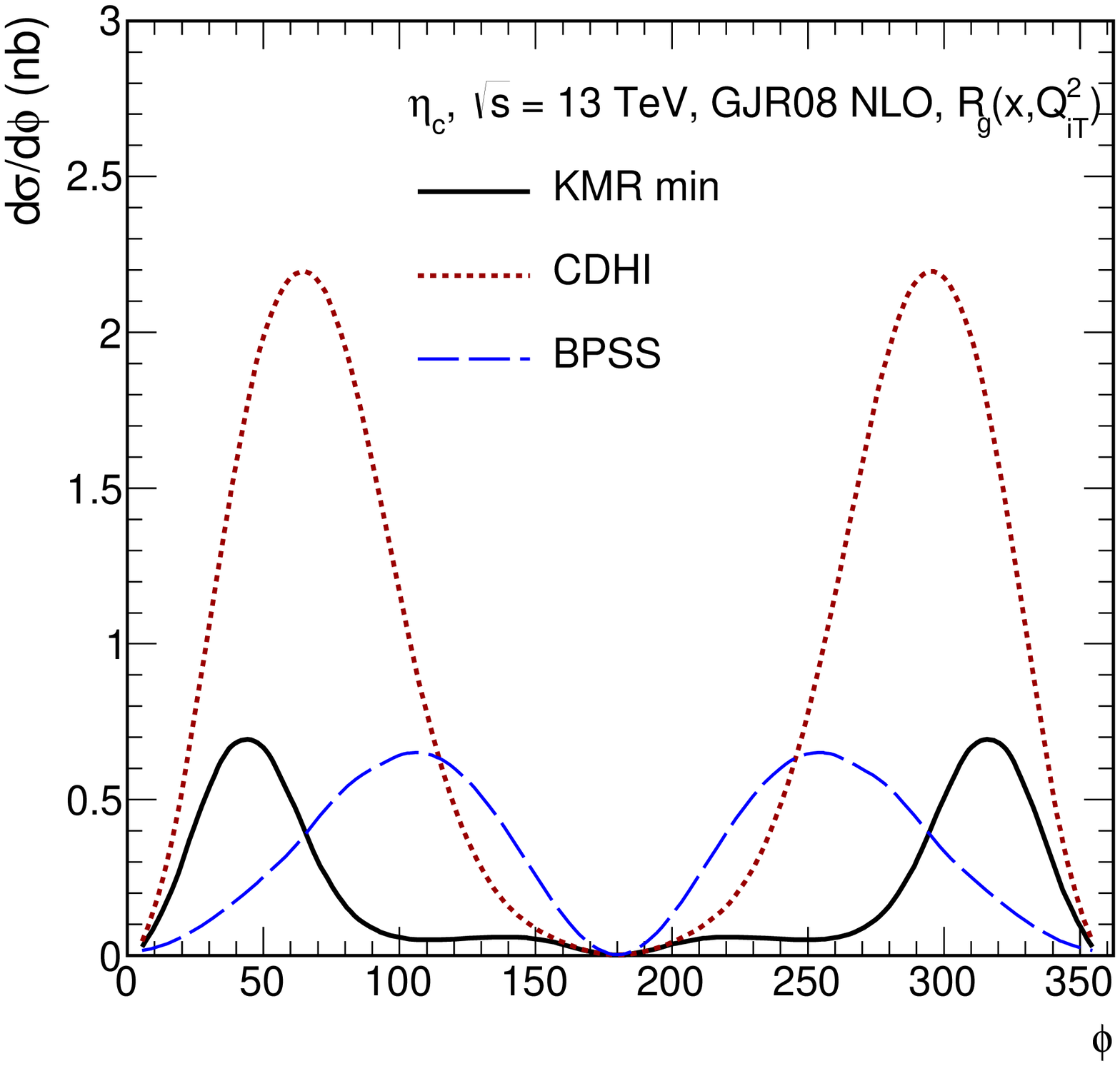}
    \caption{Distribution in relative azimuthal angle between outgoing
      protons, for $\chi_{c0}$ -- left panel, and for $\eta_c$  -- right panel, using different prescriptions for UGDs.
      No gap survival factor is included here.}
    \label{fig:dsig_dphi}
\end{figure}

In Fig.~\ref{fig:dsig_dphi} we show relative azimuthal angle (between outgoing
protons) distributions. The distribution for $\chi_{c0}$ (left)
is very different than that for $\eta_c$ (right).
While for $\chi_{c0}$ there is one maximum for the back-to-back
configurations, there are two maxima for $\eta_c$.
The cross section vanishes in the back-to-back
kinematics in the case of $\eta_c$ CEP. The exact position of the maxima depends 
on the details of the treatment of the off-diagonal UGDs so their experimental
identification could pin down the correct theoretical modelling of these objects.

\begin{figure}
    \centering
    \includegraphics[width = 0.45\textwidth]{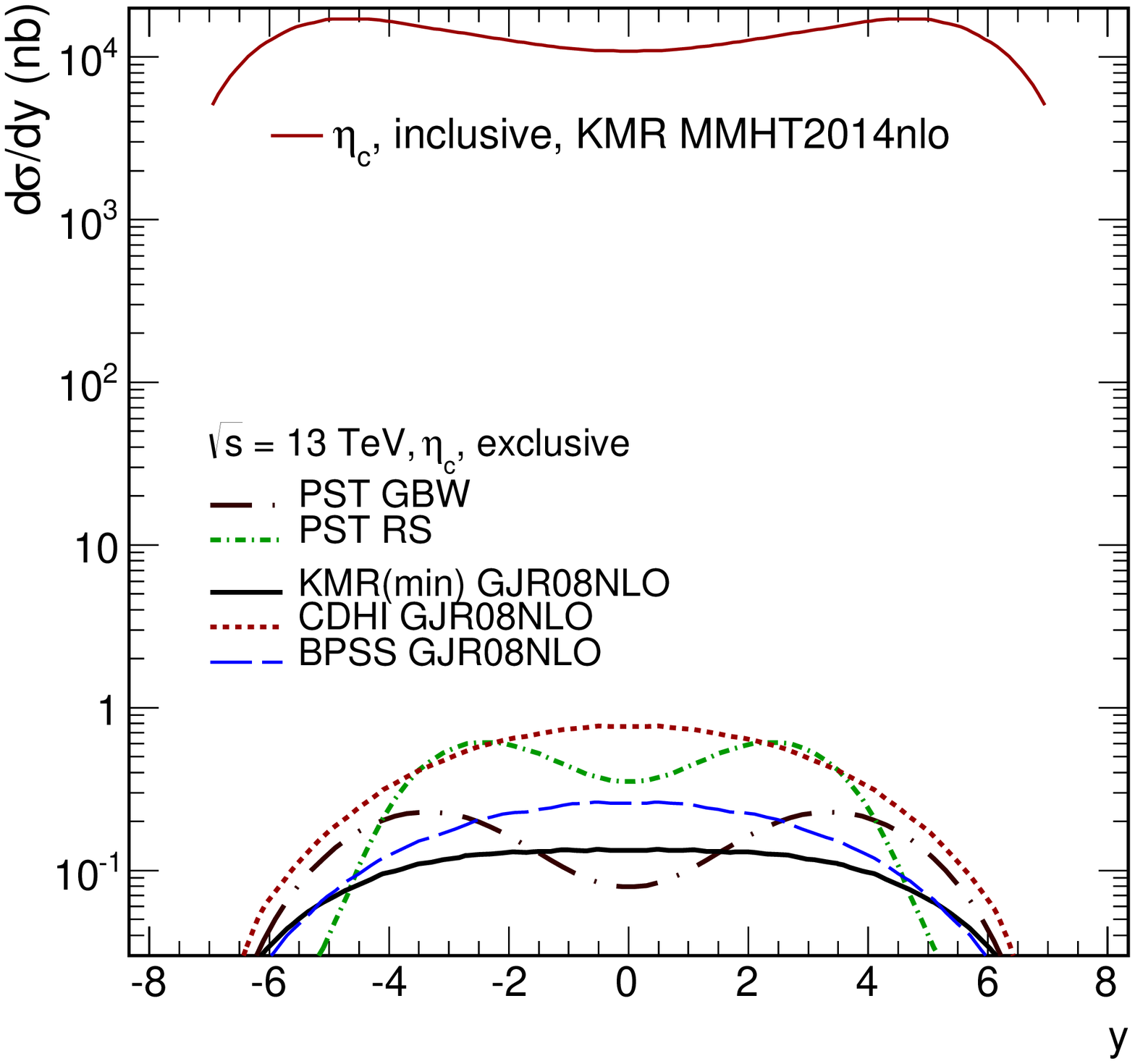}
    \includegraphics[width = 0.45\textwidth]{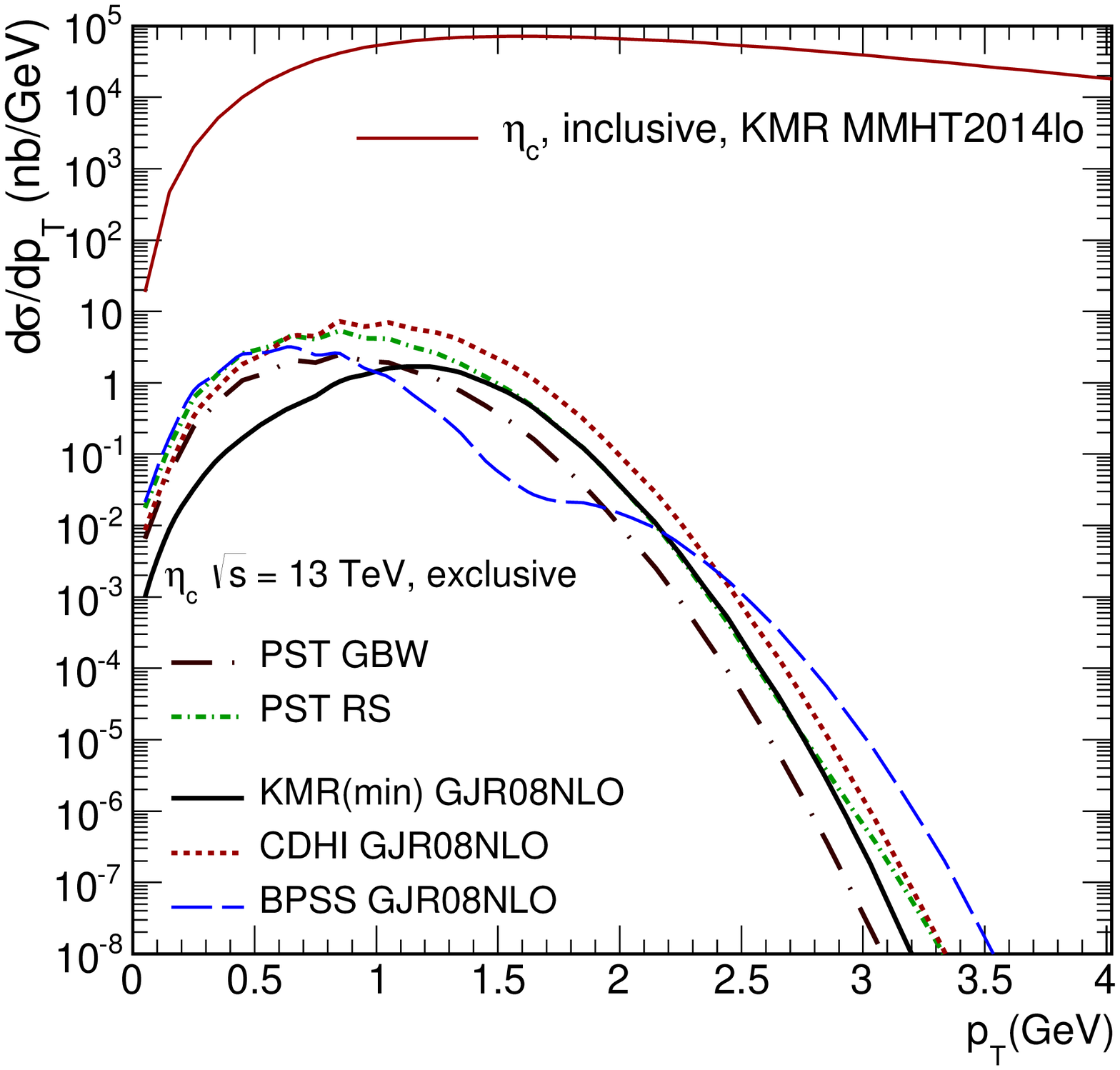}
    \caption{Comparison of exclusive and inclusive $\eta_{c}$ production
     at $\sqrt{s}= 13\,{\rm TeV}$. No gap survival factor is included
     for the exclusive reaction.}
    \label{fig:inclusive_etac}
\end{figure}

\begin{figure}
    \centering
    \includegraphics[width = 0.45\textwidth]{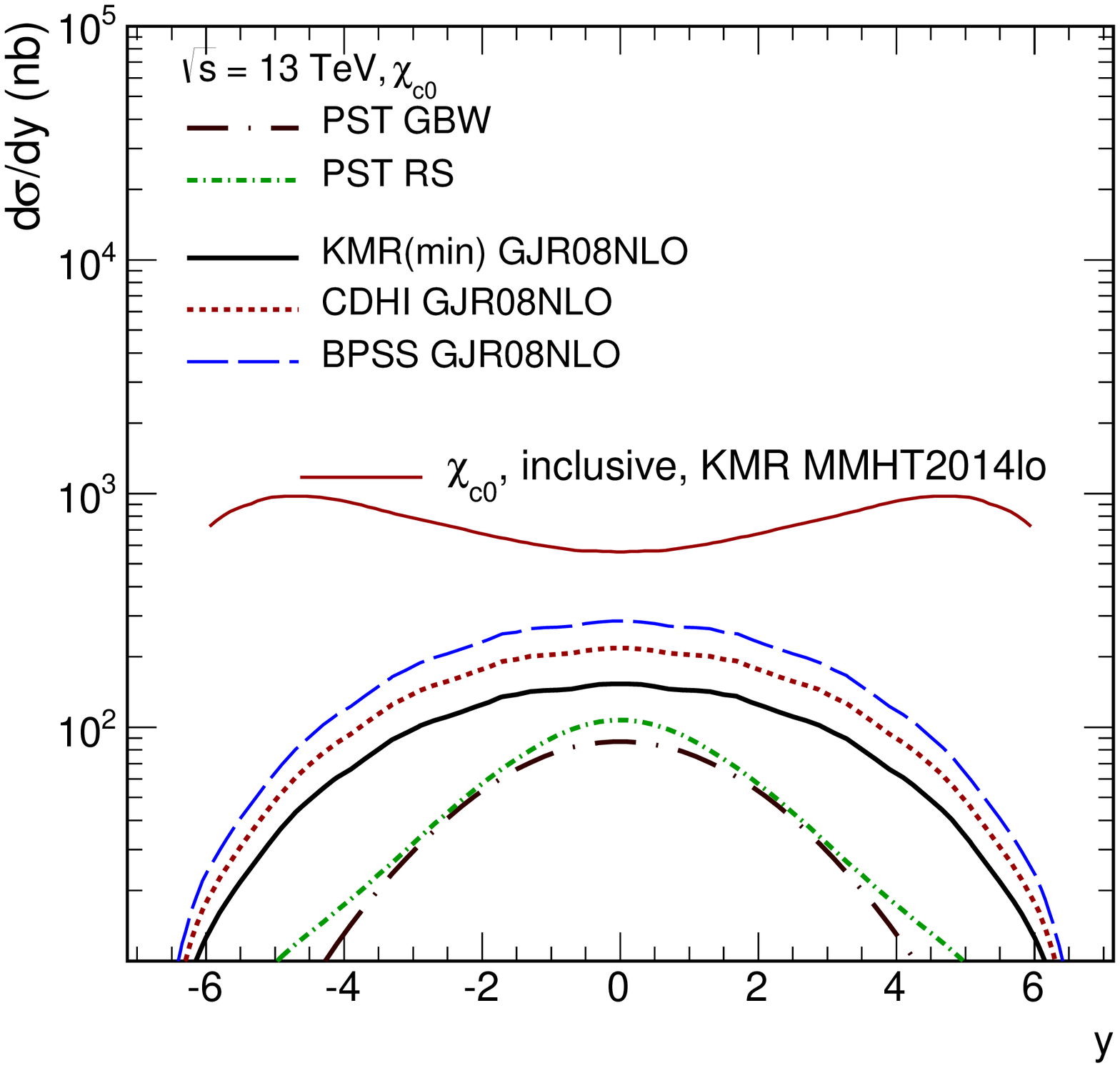}
    \includegraphics[width = 0.45\textwidth]{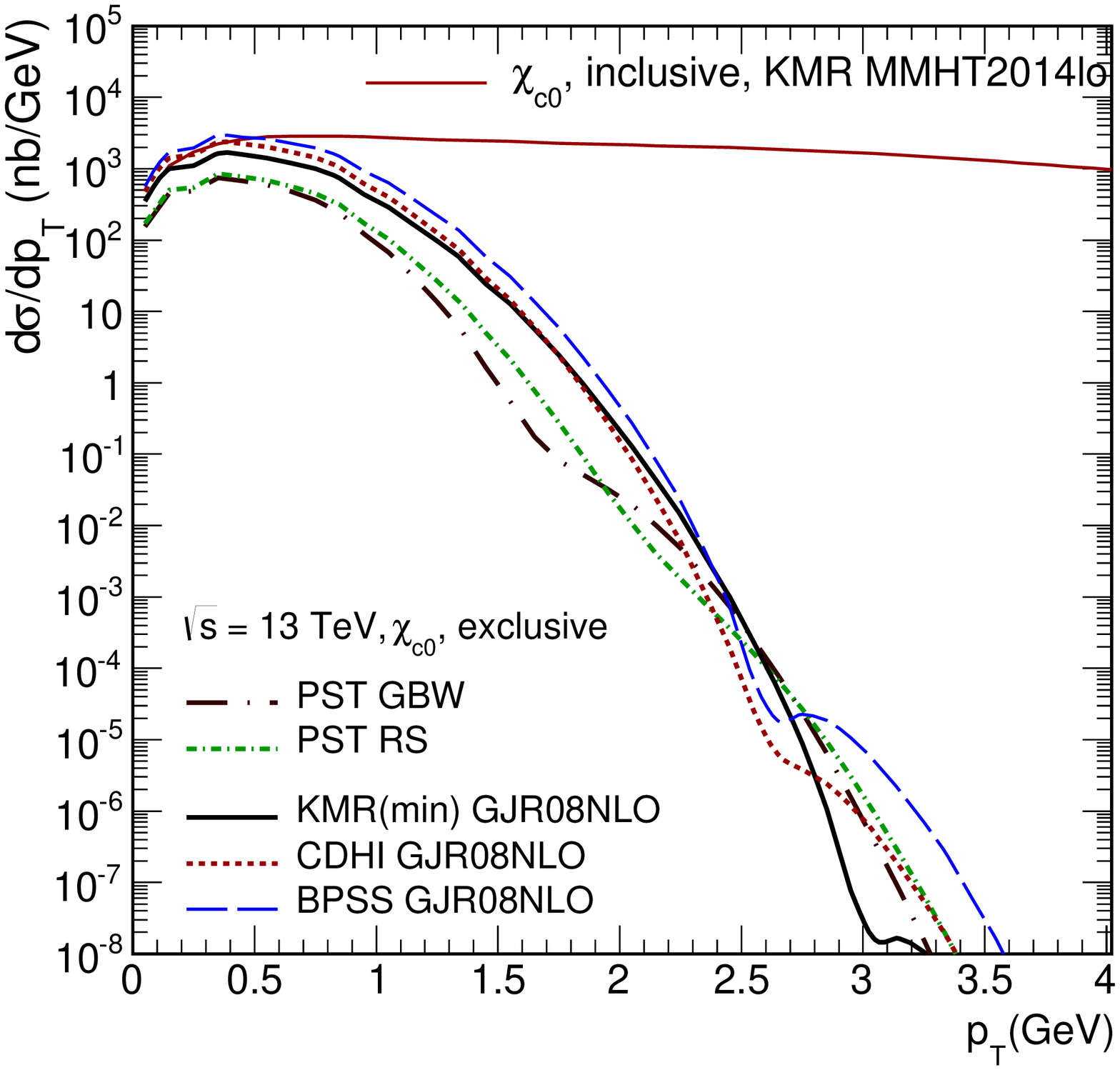}
    \caption{Comparison of exclusive and inclusive $\chi_{c0}$
      production at $\sqrt{s}= 13\,\rm{TeV}$. No gap survival factor is
      included for the exclusive reaction.}
    \label{fig:inclusive_chic0}
\end{figure}

Finally, we wish to compare our results for the exclusive reactions 
$p p \to p p \eta_c$ and $p p \to p p \chi_{c0}$ with their
inclusive production counterparts as calculated recently 
in Refs.~\cite{BPSS2019} and \cite{BPSS2020}.
In Figs.~\ref{fig:inclusive_etac} and \ref{fig:inclusive_chic0} we show
the numerical results (rapidity and transverse momentum distributions) for
$\eta_c$ and $\chi_{c0}$, respectively.
While for $\eta_c$ production the cross section for the exclusive process is a few
orders of magnitude lower than that for the inclusive case,
this is quite different for $\chi_{c0}$ meson.
Both for rapidity and transverse momentum distributions the results for
the exclusive case are very different compared to the inclusive case.

The two UGDs obtained from the dipole cross section, labelled GBW and RS, 
give rise to similar distributions. For the case of $\eta_c$ the RS UGD gives 
a larger cross section than that obtained with the GBW model, while in the case 
of $\chi_c$ their sizes are very similar.


\section{Absorptive corrections}

It is understood that the Born-level cross sections receive absorptive
corrections through hadronic rescatterings at large distances. These are related 
to the interactions of spectator partons \cite{Bjorken:1992er}. 
They give rise to the so-called gap survival probability in exclusive reactions.
The calculation of the latter poses a difficult problem, which has not been solved yet in 
a way fully consistent with the perturbative QCD approach to the production amplitude.

Numerous approaches exist in the literature, some of them are based on soft multi-Pomeron exchanges 
\cite{Gotsman:1999xq,Kaidalov:2001iz,Ostapchenko:2017prv},
while other approaches avoid the decomposition into Born term and
absorptive correction altogether treating the absorptive effects dynamically 
\cite{Flensburg:2012zy,Rasmussen:2015qgr} and at the amplitude level in 
the dipole picture \cite{Pasechnik:2011nw,Pasechnik:2012ac,Kopeliovich:2018lha}, 
and some of them relate the gap survival probability to the absence of multiparton interactions 
\cite{Lonnblad:2016hun,Babiarz:2017jxc}.

It is also understood that the gap survival must depend on the kinematics of the process. 
Here, we wish to discuss the absorptive corrections at the amplitude level, 
in a simple quantum-mechanical treatment. To this end, 
we adopt a simple effective Reggeon Field Theory motivated approach.

In the simplified case where only ``elastic rescattering'' is taken into
account, the amplitude looks as follows:
\begin{eqnarray}
{\cal M}(Y,y,\bp_1,\bp_2) = {\cal M}^{(0)}(Y,y,\bp_1, \bp_2) - \delta {\cal M}(Y,y,\bp_1,\bp_2) \, . 
\end{eqnarray}
Here $Y = \log(s/m_p^2)$ is the rapidity difference between the colliding beams 
at center-of-mass energy $\sqrt{s}$, $y$ is the cm-rapidity of the produced meson $V$, 
and $\bp_{1,2}$ are the transverse momenta of outgoing protons.

The absorptive correction is then computed as follows
\begin{eqnarray}
\delta {\cal M}(Y,y,\bp_1,\bp_2) = \int {d^2 \bk \over 2 (2 \pi)^2}
\, T(s,\bk) {\cal M}^{(0)}(Y,y,\bp_1+ \bk,\bp_2-\bk) \, , 
\end{eqnarray}
with
\begin{eqnarray}
 T(s,\bk) = \sigma^{pp}_{\rm tot}(s) \, \exp\Big(-\half B_{\rm el}(s) \bk^2 \Big) \, .
\nonumber \\
\end{eqnarray}
At $\sqrt{s} = 13 \, {\rm TeV}$ we take $\sigma^{pp}_{\rm tot} = (110.6\pm 3.4) \, {\rm mb}$ and the nuclear slope
$B_{\rm el} = (20.36\pm 0.19) \, {\rm GeV}^{-2}$ \cite{Antchev:2017dia}.
In a double-Regge approach, the Born-level amplitude has the form
\begin{eqnarray}
 {\cal M}^{(0)}(Y,y,\bp_1,\bp_2) = is \Phi_1(\bp_1) R_\Pom(Y-y, \bp_1^2) \, V(\bp_1,\bp_2) R_\Pom(y,\bp_2^2) \Phi_2(\bp_2) 
\; .
\label{eq:Regge_amplitude}
\end{eqnarray}
Here, $R_\Pom(y,\bp^2)$ are the Pomeron Regge-propagators, and $V(\bp_1,\bp_2)$ is the 
$\Pom \Pom \to {\rm Meson}$ vertex.

Let us now briefly discuss the vertices $V(\bp_1,\bp_2)$.
The most general form of the Pomeron-Pomeron-particle vertex for a spinless 
particle can be written as a Fourier expansion:
\begin{eqnarray}
V(\bp_1,\bp_2) = V_0(\bp_1^2,\bp_2^2) + \sum_{n \geq 1} \Big( V^+_n(\bp_1^2,\bp_2^2) \cos(n \phi) 
+ V^-_n(\bp_1^2,\bp_2^2) \sin(n \phi) \Big) \, .
\end{eqnarray}
For a scalar particle, all $V^-_n = 0$, while for the pseudoscalar
$V_0 = 0,\, V^+_n = 0$.
For definiteness, let us concentrate on only the first order, $n=1$.
We thus adopt ($V^{0+}$ for scalar and $V^{0-}$ for pseudoscalar state):
\begin{eqnarray}
V^{0+}(\bp_1,\bp_2) &=& V_0  + V^+_1 \, (\bp_1 \cdot \bp_2)
= V_0 \Big( 1 + \tau B_D (\bp_1 \cdot \bp_2) \Big)\, \quad  {\rm with} \, \quad \tau \equiv{ V^+_1 \over B_D V_0} \nonumber \\
V^{0-}(\bp_1,\bp_2) &=& V_1^- \, [\bp_1,\bp_2] \, .
\label{eq:approx_vertex}
\end{eqnarray}
We further neglect a possible dependence of vertices
$V_i^{\pm}$ on $\bp_1^2$ and $\bp_2^2$.
\begin{table}[]
    \centering
   \caption{$V_0$ and $\tau$ at midrapidity of $\chi_{c0}$, for several
     prescriptions for off-diagonal UGDs.}
    \begin{tabular}{l|c|c|c|c|c|c|c|c}
    \hline
    \hline
    $\chi_{c0}$ & $V_0^+$ [$\sqrt{\rm nb}/{\rm GeV}^2$] & $\tau$ & $B_D [{\rm GeV}^{-2}]$ & $g_{\rm abs}$& 
    $\beta$ & $\sigma_{\rm tot} |_{y = 0}$ [nb] & $\sigma_{\rm tot}^{\rm abs}|_{y = 0}$ [nb]& $S^2_{y=0}$\\
    \hline
    PST GBW           & $-$2062 & $-$0.31    & 5.7 & 0.71 & 0.18 & 17 & 3.7 & 0.21\\
    PST RS            & $-$2381 & $-$0.28    & 5.9 & 0.70 & 0.18 & 21 & 4.5 & 0.21\\
    CDHI GJR08NLO     & $-$2985 &  $-$0.135  & 4.5 & 0.76 & 0.15 & 42 & 7.5 & 0.18\\
    KMR  GJR08NLO     & $-$2167  & $-$0.11   & 4.5 & 0.77 & 0.15 & 29 & 3.7 & 0.13\\
    BPSS GJR08NLO     & $-$3118  & $-$0.135  & 4.5 & 0.77 & 0.15 & 61 & 8.0 & 0.13\\
    \hline
        \end{tabular}
    \label{tab:V0V1}
\end{table}
\begin{figure}
    \centering
    \includegraphics[width = 0.3 \textwidth]{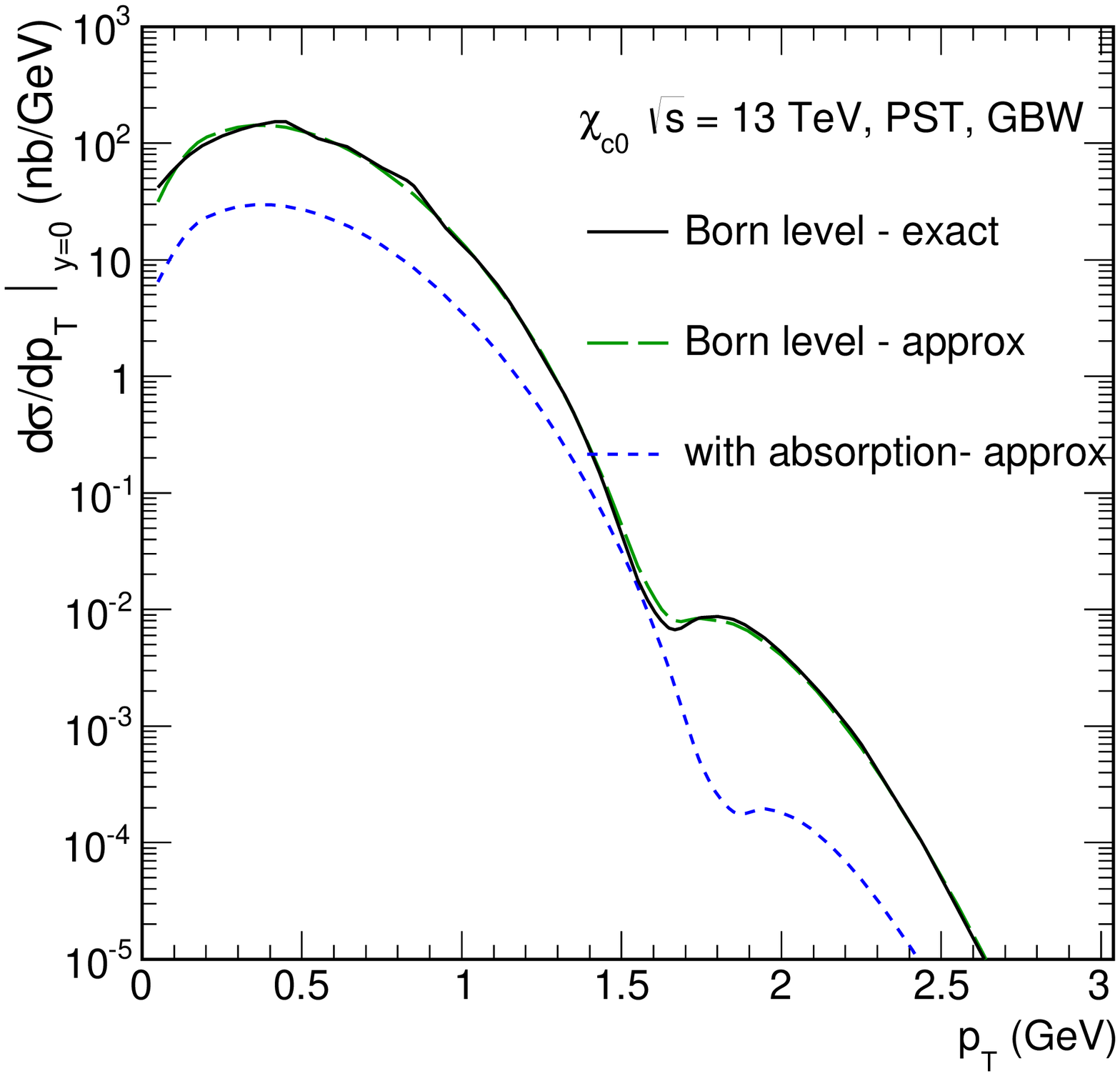}
     \includegraphics[width = 0.3 \textwidth]{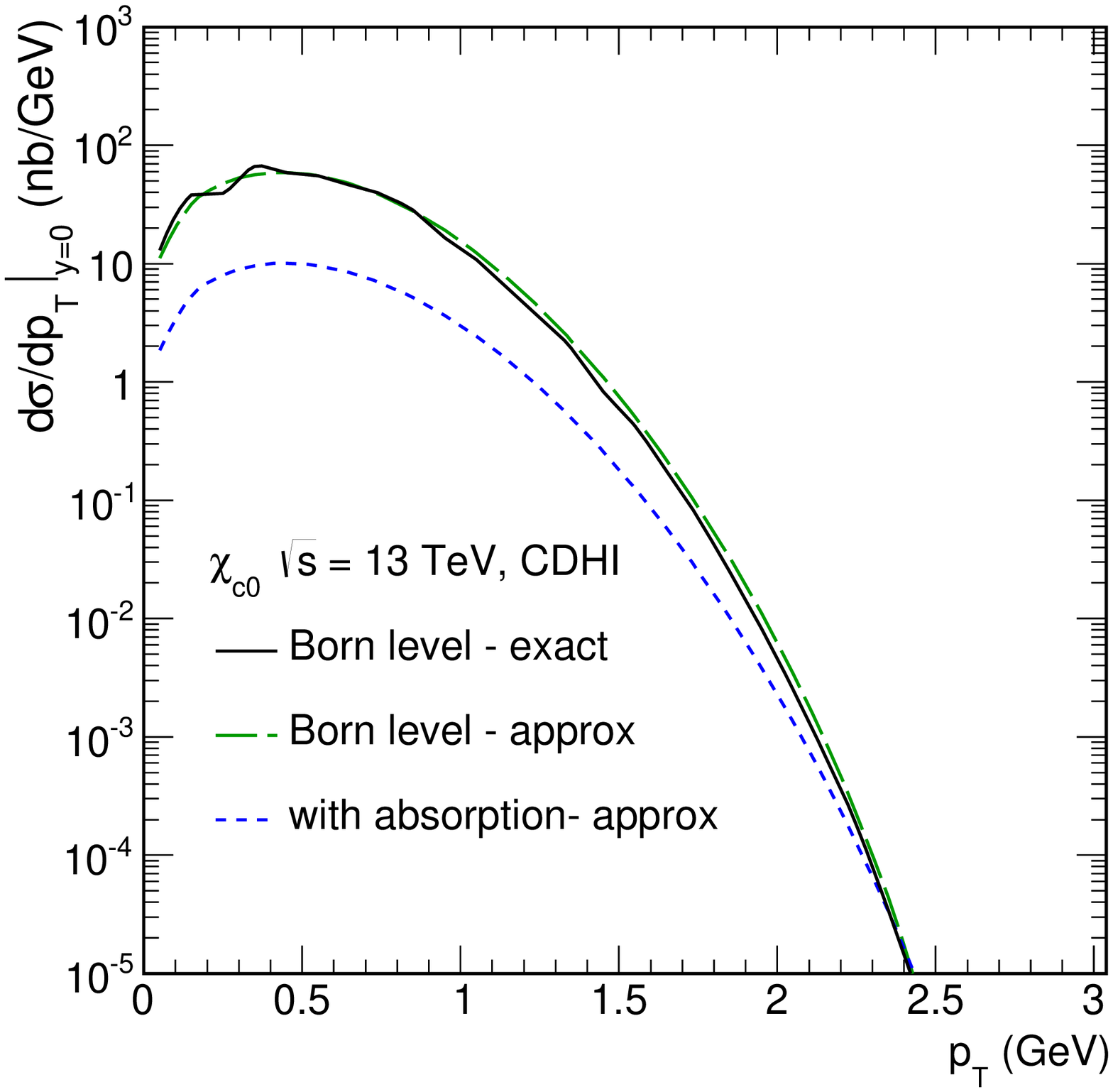}
     \includegraphics[width = 0.3 \textwidth]{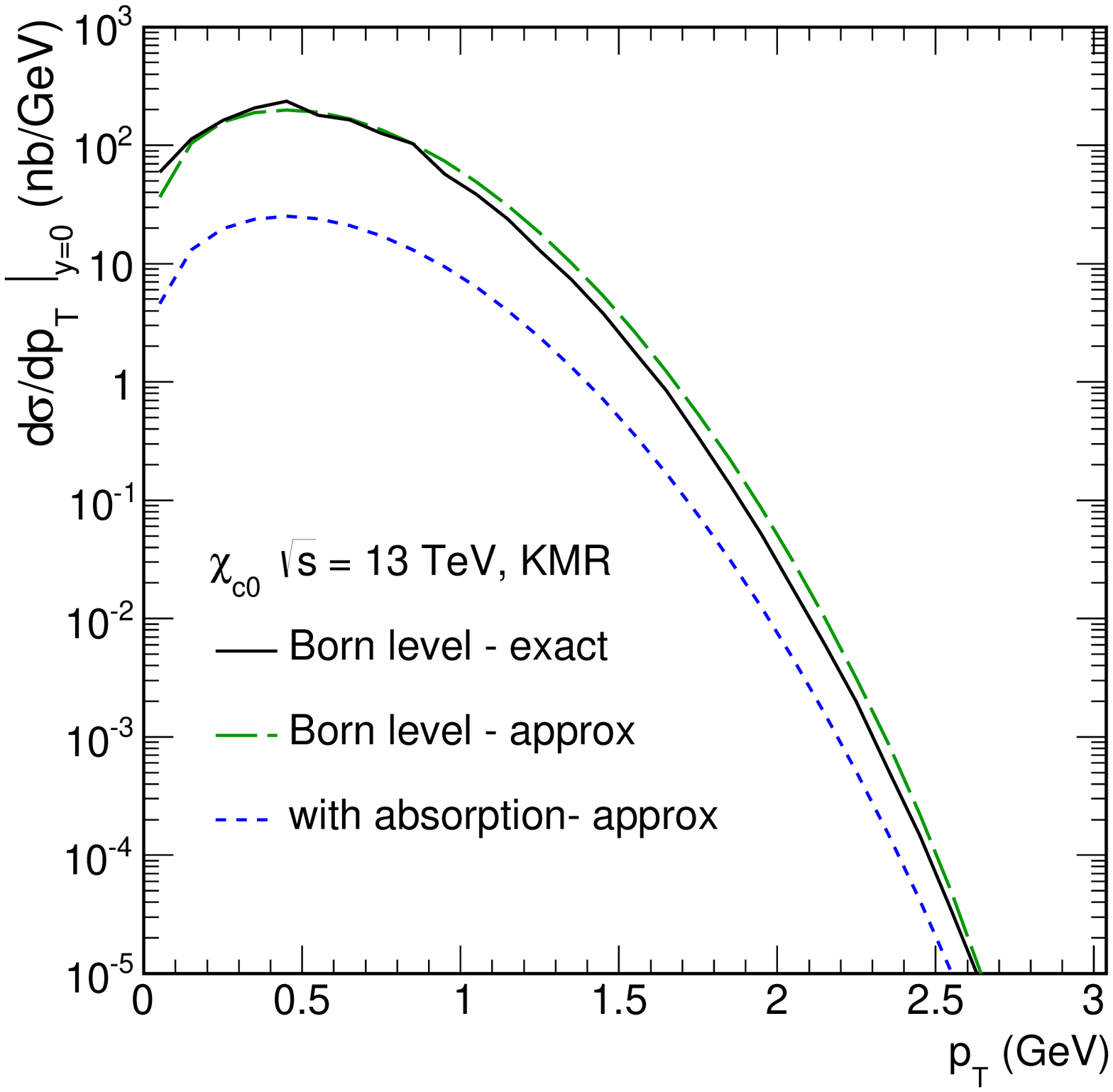}
    \caption{Transverse momentum distribution of $\chi_{c0}$ CEP at $y=0$, with the
      PST-GBW, the CDHI and Durham prescription.}
   \label{fig:dsig_dpt_chic_y0}
\end{figure}

\begin{table}[]
    \centering
   \caption{An example of $V_1$ values at midrapidity of $\eta_{c}$ in the CEP process,
     for several prescriptions for off-diagonal UGDs.}
    \begin{tabular}{l|c|c|c|c|c|c|c}
         \hline
         \hline
     $\eta_c$ & $V_1^-$ [$\sqrt{\rm nb}/{\rm GeV}^4$]  & $B_D [{\rm GeV}^{-2}]$  & $g_{\rm abs}$ & 
     $\beta$ & $\sigma_{\rm tot}|_{y = 0}$ [nb] & $\sigma_{\rm tot}^{\rm abs}|_{y = 0}$ [nb] & $S^2_{y=0}$\\
     \hline
    PST GBW       & 194.    & 3.4 & 0.83 & 0.12 & $1.8\times10^{-2}$ & $3.9\times10^{-3}$ & 0.21\\
    PST RS        & 400.    & 3.2 & 0.84 & 0.12 & $9.0\times10^{-3}$ & $1.9\times10^{-3}$ & 0.21\\
    CDHI GJR08NLO & 651.   & 3.5 & 0.81 & 0.13 & $1.8\times10^{-1}$ & $4.0\times10^{-2}$ & 0.22 \\
    KMR  GJR08NLO & 1015.   & 4.7 & 0.76 & 0.16 & $1.3\times10^{-1}$ & $3.0\times10^{-2}$ & 0.29\\
    BPSS GJR08NLO & 1490.   & 7.0 & 0.66 & 0.20 & $5.8\times10^{-2}$ & $2.2\times10^{-2}$ & 0.38\\
    \hline
    \end{tabular}
    \label{tab:V0V1_etac}
\end{table}
\begin{figure}
    \centering
    \includegraphics[width = 0.3\textwidth]{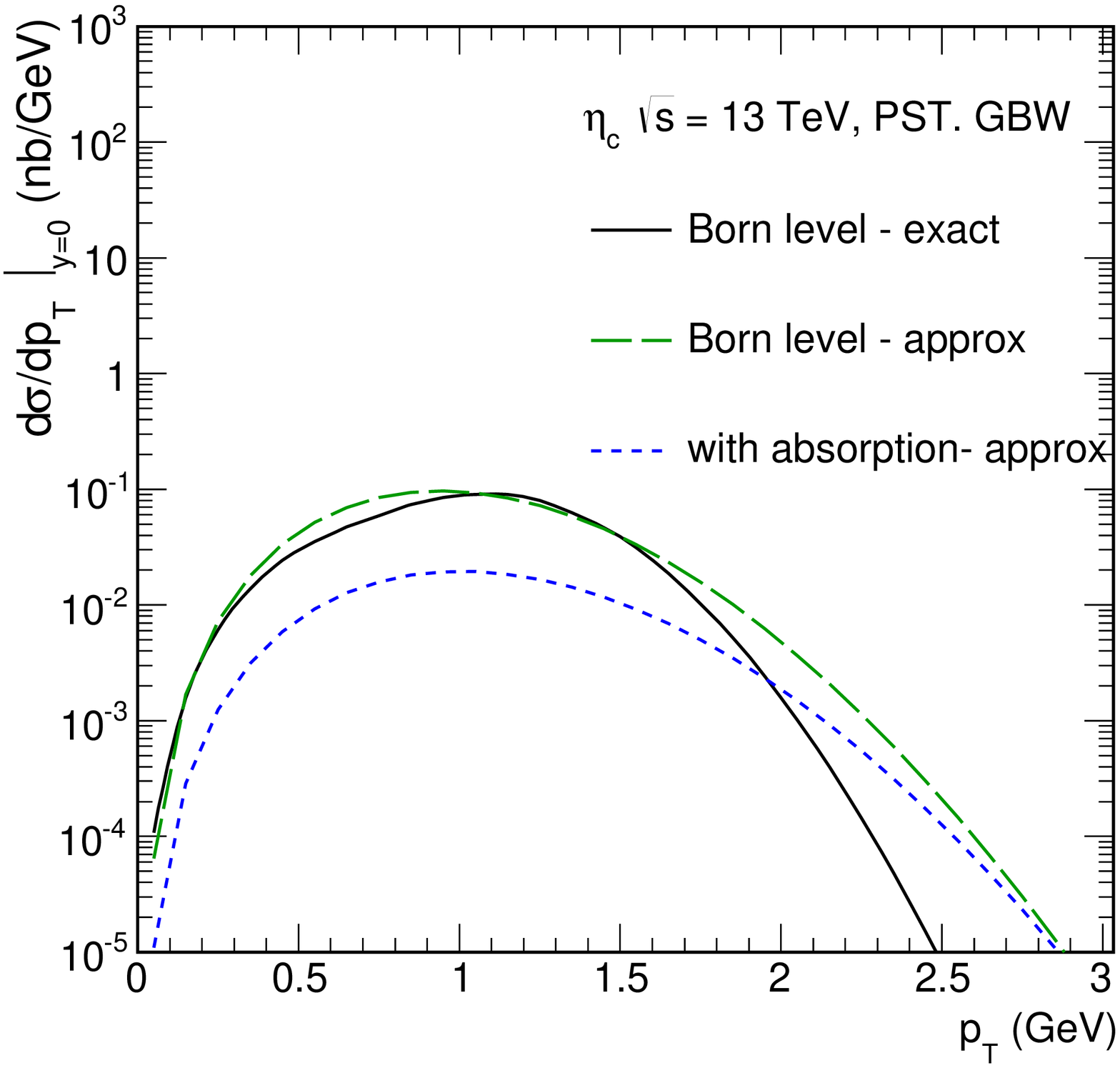}
    \includegraphics[width = 0.3\textwidth]{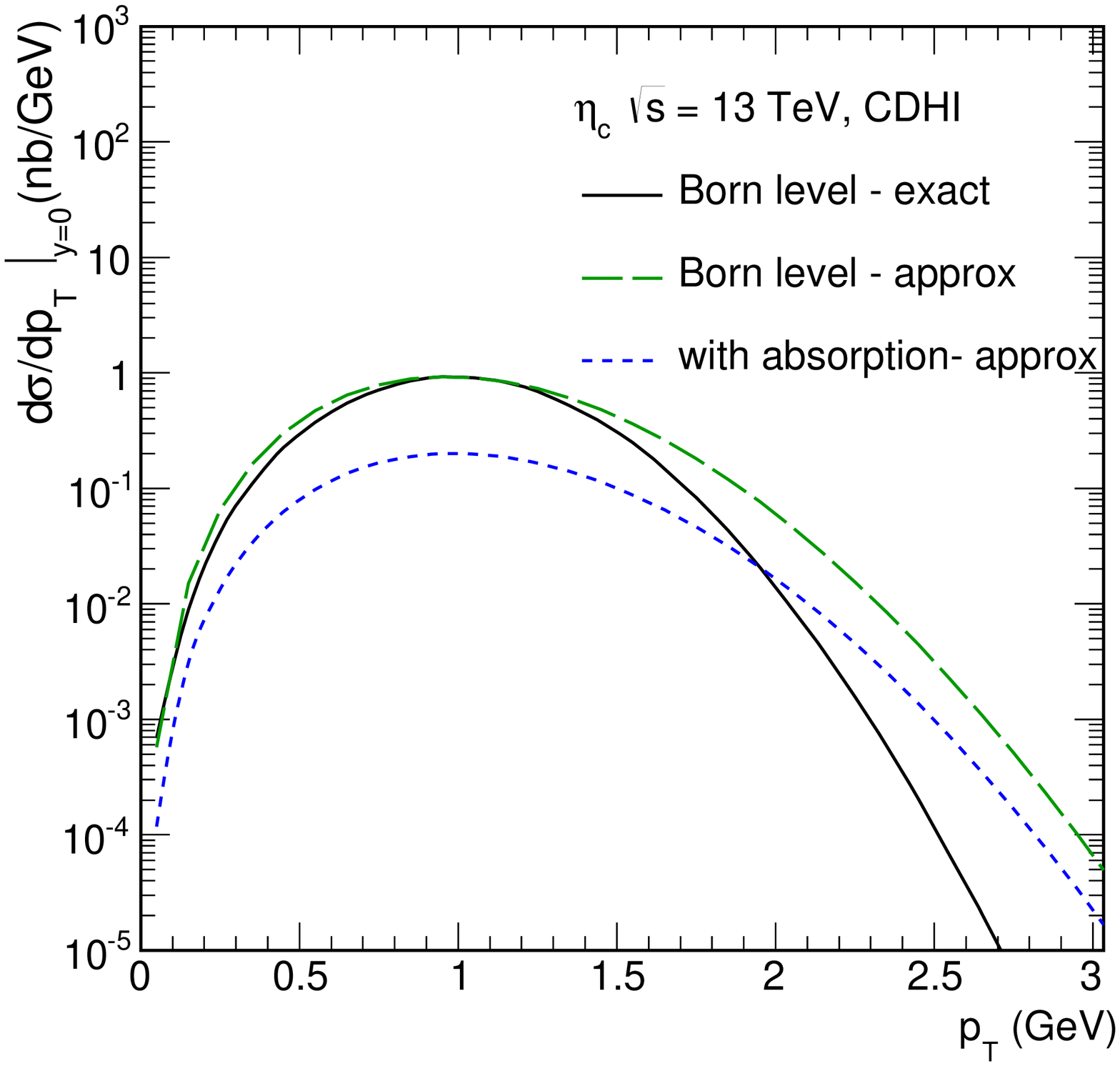}
    \includegraphics[width = 0.3 \textwidth]{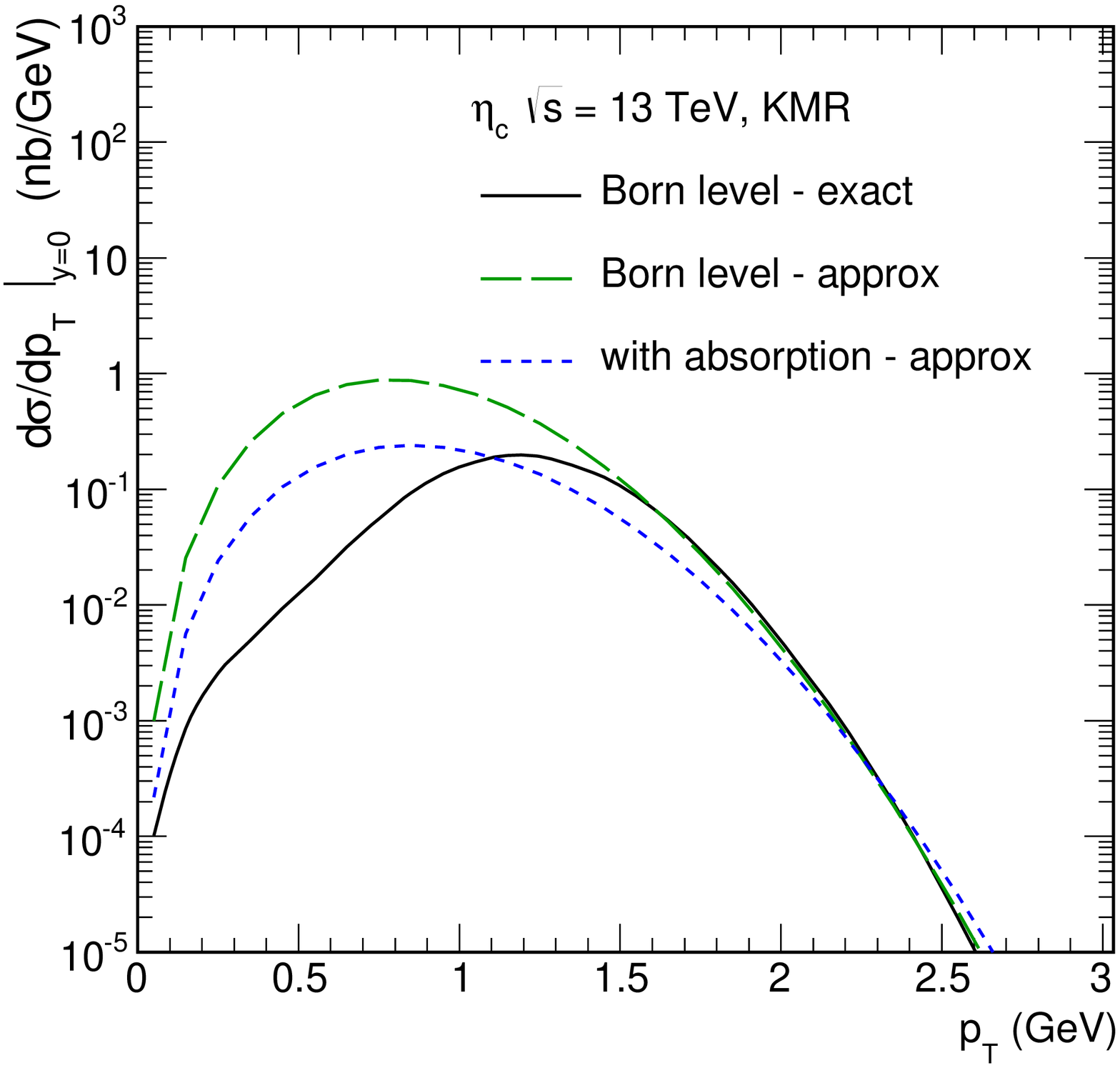}
    \caption{Distribution in transverse momentum of $\eta_c$ in the CEP process for
      PST-GBW, CDHI and Durham prescriptions at $y = 0$.}
    \label{fig:dsig_dpt_etac_y0}
\end{figure}

Our amplitude is normalized in such a way that the expression
\begin{eqnarray}
d \sigma = {1 \over 256 \pi^5 s^2} |{\cal M}(Y,y,\bp_1,\bp_2)|^2 dy d^2\bp_1 d^2\bp_2 d^2\bp \, 
\delta^{(2)}(\bp + \bp_1 + \bp_2)
\end{eqnarray}
holds. We will now concentrate on central diffractive production, i.e. 
we fix the meson rapidity to be $y=0$.
Below we adopt $\sqrt{s}= 13 \, {\rm TeV}$.
We can therefore forget about the Regge-propagators in
Eq.~(\ref{eq:Regge_amplitude}), and without loss of generality we write
\begin{eqnarray}
\Phi_{1,2}(\bp_{1,2}) = \exp\Big( -\half B_D \bp^2_{1,2} \Big) \, .
\end{eqnarray}

Then, using the vertices of Eq.~(\ref{eq:approx_vertex}), the transverse momentum distributions of the mesons
at the Born level are obtained as 
\begin{eqnarray}
{d \sigma^{0+}_{\rm Born} \over dy  d p^2_T}\Big|_{y=0} &=& 
{\exp[- \half B_D p_T^2] V_0^2 \over 512 \pi^3 B_D}
\Big\{
1  -  \tau (1 - \half B_D p_T^2)   
+  {\tau^2 \over 2} \Big( 1 - \half B_D p_T^2 + {1 \over 8} B_D^2 p_T^4 \Big) \Big\} \nonumber \\
{d \sigma^{0-}_{\rm Born} \over dy  dp_T^2}\Big|_{y=0} &=&
{(V_1^-)^2 \over 512 \pi^3} {p_T^2 \over 4B_D^2} \exp[-\half B_D p_T^2] \, .
\end{eqnarray}
Now, the absorptive corrections require the evaluation of the loop integral
\begin{eqnarray}
\delta {\cal M}(Y,0,\bp_1,\bp_2) &=& \int {d^2 \bk \over 2 (2 \pi)^2}
\, T(s,\bk)  \exp\Big( -\half B_D (\bp_1 + \bk)^2 \Big)\exp\Big( -\half B_D (\bp_2-\bk)^2 \Big) \nonumber \\
&\times&  V(\bp_1+ \bk,\bp_2-\bk)  
=\exp\Big(-\half B_D( \bp_1^2 + \bp_2^2) \Big)
\nonumber \\
&\times& \int {d^2 \bk \over 2 (2 \pi)^2}
\exp\Big( - \half(B_{\rm el}(s) + 2 B_D) \bk^2 -B_D \bk \cdot (\bp_1 - \bp_2)  \Big) \, \nonumber \\
&\times& {\sigma^{pp}_{\rm tot}(s) \, V(\bp_1+ \bk,\bp_2-\bk)} \, .
\end{eqnarray}
It is useful to introduce the dimensionless quantities
\begin{eqnarray}
g_{\rm abs} = {\sigma^{pp}_{\rm tot}(s) \over 4 \pi( B_{\rm el}(s) + 2 B_D)} \, \quad 
{\rm and} \quad \, \beta = {B_D \over B_{\rm el}(s) + 2 B_D} \, .
\end{eqnarray}
Then, the absorptive corrections are obtained as
\begin{eqnarray}
\delta {\cal M}^{0+}(Y,0,\bp_1,\bp_2) &=& 
g_{\rm abs} V_0 \exp\Big( - \half B_D (\bp_1^2 + \bp_2^2) \Big) \, 
\exp\Big( {1 \over 2} \beta B_D (\bp_1 - \bp_2)^2 \Big) 
\nonumber \\
&\times& \Big \{ 1 + \beta(1+\beta) \tau B_D  (\bp_1^2 + \bp_2^2)  
+ (\bp_1 \cdot \bp_2) \tau B_D \Big( 1 - 2 \beta(1+\beta) \Big) \Big\} \,,
\nonumber \\
\end{eqnarray}
\begin{eqnarray}
\delta {\cal M}^{0-}(Y,0,\bp_1,\bp_2) &=& 
(1-\beta) g_{\rm abs} V_1^- 
\exp\Big( - \half B_D (\bp_1^2 + \bp_2^2) \Big) \, \exp\Big( {1 \over 2} \beta B_D (\bp_1 - \bp_2)^2 \Big) 
\nonumber \\
&\times& [\bp_1,\bp_2]
(1-\beta) \Big(1 -\beta B_D (\bp_1 \cdot \bp_2) \Big) \, ,
\end{eqnarray}
for the scalar and pseudoscalar meson, respectively.
We now adjust the constants $V_0, V_1^\pm$,
as well as $B_D$, to our numerical results obtained 
for the Born-level amplitude. 

In Tables~\ref{tab:V0V1} and \ref{tab:V0V1_etac}, we show the parameters 
obtained for different prescriptions for the generalized unintegrated gluon
distribution, GBW as well as the CDHI and Durham prescriptions for 
the GJR08NLO gluon distribution. We also show the gap survival factors
\begin{eqnarray}
S^2 \equiv  \frac{d \sigma/dy \Big|_{y=0}} {d \sigma_{\rm Born}/  dy \Big|_{y=0}} \, .
\end{eqnarray}
We observe that depending on the gluon distribution used, we obtain 
for the $\chi_c$ the gap survival values of $S^2 = 0.13 \div 0.21$, 
while for the $\eta_c$ production they are systematically somewhat higher,  $S^2 = 0.21 \div 0.38$. Notice, that the $\eta_c$ amplitude, due to the vanishing in forward direction, is more peripheral than the one for $\chi_c$ production. 
However also notice that the both reactions have significantly different values of the effective diffraction slopes $B_D$. Regarding
the diffraction slope we furthermore observe a strong model dependence, especially for the $\eta_c$ case. 

Our simplified double-Regge approach works reasonably well. 
In Fig.~\ref{fig:dsig_dpt_chic_y0}, we show
the cross section $d\sigma/dy dp_T$ at $y=0$ 
for the $\chi_c$ for the three different generalized UGD prescriptions. 
Shown is the exact numerical result of the Born amplitude (solid line)
as well as the result of our effective Regge amplitude 
fit (long-dashed line).
By the short-dashed line we show the differential cross section
including absorptive corrections on top of 
the Regge amplitude Born term. We see from these figures that the
effective Regge amplitude form 
is reasonably accurate for $p_T \lsim 1.5 \, {\rm GeV}$, with 
a slight ambiguity in the slope $B_D$.
In the case of the $\eta_c$ shown in Fig.~\ref{fig:dsig_dpt_etac_y0},
the effective Regge fit works almost perfectly for the PST-GBW and CDHI
prescriptions, while for the Durham case the description is rather poor. 

In our calculation of absorptive corrections, we restricted ourselves to the so-called elastic rescattering correction. We wish to point out, that the often applied multichannel models
that account for the possible diffractively excitated intermediate states, are constructed for
soft diffractive processes. 
In our case we deal with a Born level processes
with (semi-)hard gluon exchanges, which will favour a coupling to small color dipoles in each proton. It is not clear that the diffractive final
states that dominate soft diffractive dissociation at the LHC have a large overlap
with the relevant dipole sizes. 

In this regard we wish to point to the work of Refs. \cite{Pasechnik:2011nw,Pasechnik:2012ac,Kopeliovich:2018lha}. In these works it is argued, that for certain hard
inclusive diffractive processes (part of) the rescattering corrections are in fact already included effectively in the dipole cross section.
The consistent formalism for exclusive channels remains an important task for the future. 

\section{Conclusion}

In the present paper we have calculated the key observables
of central exclusive $\chi_{c0}$ and $\eta_c$ quarkonia production
in proton-proton collisions at the LHC within 
a formalism proposed earlier by the Durham group 
for central exclusive Higgs boson production.

The $\chi_{c0}$ meson CEP was already computed in the literature 
previously, while $\eta_c$ production has been analysed here for the first time.
Compared to the previous calculations we have used here 
modern versions of collinear gluon distributions to generate
off-diagonal unintegrated gluon distributions.

In the present analysis we have also used the $g g \to \eta_c$
and $g g \to \chi_{c0}$ transition amplitudes calculated using the light-cone $c \bar c$ 
wave functions obtained in the framework of potential models.
We have performed similar calculations for inclusive production of $\eta_c$ and
$\chi_{c0}$ very recently and showed that one can very well describe the
experimental data for $\eta_c(1S)$ meson measured in last few years by 
the LHCb collaboration. Our previous results showed that in the inclusive 
case the cross section for $\eta_c$ is significantly larger 
than that for $\chi_{c0}$.

It was the main aim of the present paper to make a similar analysis for the exclusive production 
case, which for the case of $\eta_c$ not done so far in the literature. In contrast to the inclusive case, we have found 
that for the CEP the situation reverses i.e. the corresponding cross section for exclusive
$\eta_c$ production is considerably smaller than its counterpart for exclusive $\chi_{c0}$ 
production, at least, for the hard part obtained using the Durham or Cudell et al 
prescriptions for calculation of the scale in the off-diagonal unintegrated gluon 
distribution. The reason is a specific interplay of the off-diagonal UGDs and
virtual gluon -- virtual gluon -- quarkonium vertex.

We also proposed a way to calculate the soft effects (in the region of small gluon
transverse momenta) using the GBW or RS UGDs, which were obtained from the respective color dipole cross sections,  and a simple (PST) prescription for its off-diagonal 
extrapolation. In this case, the cross section is only slightly smaller for $\eta_c$ than 
for $\chi_{c0}$ production. We have also discussed to which extent the absorption effects for
$p p \to p p \eta_c$ are different than those for $p p \to p p \chi_{c0}$.
We find that the absorptive corrections for the
$\eta_c$ somewhat smaller, which correlates with a  very different $(t_1,t_2)$ dependence for the corresponding Born amplitudes. 
However, there is a rather strong model dependence on the Born-amplitude. 

It would be desirable to measure the cross section for $p p \to p p \eta_c$
by identifying $\eta_c$ e.g. in the $p \bar p$ decay channel as was done
in the inclusive case. It could be interesting to
estimate the signal-to-background ratio before the real experiment.
The $p p \to p p p \bar p$ continuum was calculated previously by Lebiedowicz, Nachtmann
and Szczurek \cite{LNS_ppbar} and a first experimental evidence was
obtained very recently by the STAR collaboration at RHIC \cite{STAR_exclusive}.
Also the $p p \to p p \gamma \gamma$ reaction could be considered as an
alternative to measure the $p p \to p p \eta_c$ reaction.

\section*{Acknowledgments}
The stay of I.B. in Lund was supported by Polish National Agency for Academic Exchange
under Contract No. PPN/IWA/2018/1/00031/U/0001.
This study was partially supported by the Polish National Science Center
under grant No. 2018/31/B/ST2/03537 and by the Center for Innovation
and Transfer of Natural Sciences and Engineeing Knowledge in Rzesz\'ow (Poland).
R.P. was partially supported by the Swedish Research Council grant No.~2016-05996 and 
by the European Research Council (ERC) under the European Union's Horizon 2020 
research and innovation programme (grant agreement No 668679).


\end{document}